Neurotrophic Effects of Intermittent Fasting, Calorie Restriction and Exercise: A Review and Annotated Bibliography

Eric Mayor,
University of Basel, Switzerland

Corresponding author:
Eric Mayor
University of Basel,
Missiosstrasse 62A
4055 Basel, Switzerland

ericmarcel.mayor@unibas.ch

Abstract.

In the last decades, important progress has been achieved in the understanding of the neurotrophic effects of intermittent fasting (IF), caloric restriction (CR) and exercise. Improved neuroprotection, synaptic plasticity and adult neurogenesis (NSPAN) are essential examples of these neurotrophic effects. The importance in this respect of the metabolic switch from glucose to ketone bodies as cellular fuel has been highlighted. More recently, calorie restriction mimetics (CRMs; resveratrol and other polyphenols in particular) have been investigated thoroughly in relation to NSPAN. In the narrative review sections of this manuscript, recent findings on these essential functions are synthesized and the most important molecules involved are presented. The most researched signaling pathways (PI3K, Akt, mTOR, AMPK, GSK3β, ULK, MAPK, PGC-1α, NF-κB, sirtuins, Notch, Sonic hedgehog and Wnt) and processes (e.g., anti-inflammation, autophagy, apoptosis) that support or thwart neuroprotection, synaptic plasticity and neurogenesis are then briefly presented. This provides an accessible entry point to the literature. In the annotated bibliography section of this contribution, brief summaries are provided of about 30 literature reviews relating to the neurotrophic effects of interest in relation to IF, CR, CRMs and exercise. Most of the selected reviews address these essential functions from the perspective of healthier aging (sometimes discussing epigenetic factors) and the reduction of the risk for neurodegenerative diseases (Alzheimer's disease, Huntington's disease, Parkinson's disease) and depression or the improvement of cognitive function.

Neurotrophic effects of intermittent fasting, calorie restriction and exercise: A narrative review and annotated bibliography

Exercise, intermittent fasting and calorie restriction (EIC) have common beneficial effects on energy metabolism and signaling pathways: All three can induce metabolic switching from glucose to ketone bodies (g–to–k) as cellular fuel source when glucose is scarce (Mattson et al., 2018). By lowering of available cellular energy and through the cellular sensing of limited availability of nutrients[1] which ensues, they can induce the activation or inhibition of several signaling pathways relevant for neuroprotection, synaptic plasticity and neurogenesis (NSPAN) (e.g., García–Rodríguez & Giménez–Cassina, 2021; Mattson, 2019; Van Praag et al., 2014). This can have short-term implications for cognitive performance (e.g., Diano et al., 2006). The benefits of EIC are achieved in part through the influence of hunger hormone ghrelin on glucose homeostasis and the activity of the insulin receptor (Bayliss et al., 2016; Chabot et al., 2014) which is an upstream effector of signaling pathways relevant for NSPAN (Sadria & Layton, 2021). The hormone ghrelin also plays a role in other signaling pathways regulating NSPAN (Bayliss et al., 2016; Buntwal et al., 2019; Davies, 2022; Diano et al., 2006; Gahette et al., 2011). Exerkines can also regulate the activity of these pathways among others (e.g., Chow et al., 2022; Reddy et al., 2022; Vints et al., 2022). Calorie restriction mimetics (CRMs; polyphenols and other sirtuin activators as well as precursors of nicotinamide adenine dinucleotide, in particular) do not rely upon glucose depletion, but they can produce similar effects, including enhanced energy metabolism, decreased oxidative stress and inflammation and enhanced NSPAN and cognition (Testa et al., 2014). The efficacy of CRMs with regards to overall physiological parameters has been questioned by Handschin (2016), but general consensus in the literature points out to the potential of CRMs in the prevention and treatment of neurodegenerative disorders (e.g., Bonkowski & Sinclair, 2016). CRMs are considered promising in this area (e.g., Almendariz–Palacios et al., 2020; Bonkowski & Sinclair, 2016; Carosi & Sargeant, 2019; Heberden, 2016; Hofer et al., 2021; Martel et al., 2021; Sharman et al., 2019; Xue et al., 2021), but the state of current knowledge is insufficient to recommend CRMs for the treatment of individuals presenting with neurodegenerative disorders. For instance, Heger (2017) mentions that only 17 of 45 trials, at that time, show an efficacy of curcumin (in relation to diverse clinical targets). Hofer et al., (2021) provide a summary of human clinical trials interested in various CRMs relating to a diversity of outcomes. Martel et al., (2021) focus on human clinical trials interested in CRMs and the extension of human healthy lifespan, including neuroprotection and neurogenesis, discussing advantages and disadvantages. Treatment should always be provided after sufficient clinical trials have been conducted, but early in the progression of the disease (e.g., Carosi & Sargeant, 2019; Sharman et al., 2019).

EIC and CRMs are beneficial in the attenuation of epigenetic changes induced by aging, which are important contributors to neurodegeneration among other risks (Almendariz–Palacios et al., 2020; Madeo et al., 2019; Maharajan et al., 2020; Mattson et al., 2018; Xue et al., 2021). The aims of this article are threefold: A) to discuss the relevance of EIC and CRMs to NSPAN relying upon recent publications, B) to briefly present some of the biochemical processes involved in metabolism, oxidative stress and NSPAN and C) to present a selective annotated bibliography of existing literature reviews (summaries in Appendix) that focus on the ways in which EIC and CRMs can improve NSPAN, cognitive function, and brain resilience (e.g., van Praag et al., 2014).

---

[1] E.g.: increased adenosine monophosphate (AMP) to adenosine triphosphate (ATP) ratios sensed by the 5' adenosine monophosphate-activated protein kinase (AMPK) (Hardie, 2018)

**A) The Relevance of Exercise, Calorie Restriction (CR) and Intermittent Fasting (IF) and CRMs in Neurogenesis and Neuroprotection and Other Neurotrophic Effects:**
The body can transition from using glucose for energy to using ketone bodies (García-Rodríguez & Giménez-Cassina, 2021; for more details, see section B below): The liver glycogen store can contain up to the equivalent of about 700 calories. It is depleted when energy intake doesn't occur for a period of 10 to 12 hours at least, or during periods of high energy expenditure (e.g., running for one hour following a four hours period after the last meal) (Mattson et al., 2018) as glycogen is also stored in the skeletal muscles (Richter & Hargreaves, 2013). EIC can result in the depletion of the glycogen store and thus can activate the g-to-k metabolic switch. The k-to-g metabolic switch occurs when the glycogen store is repleted, after energy intake. The repeated switching between glucose and ketone bodies as main sources of cellular fuel has advantages at the organism level in terms of cellular regeneration and protection, including in the brain in terms of neurogenesis and neuroprotection (Matteson et al., 2017). CRMs such as resveratrol and rapamycin have potential in the prevention and treatment of neurodegenerative diseases (Albani et al., 2010; Baur & Sinclair, 2006; dos Santos et al., 2022; Rege et al., 2014). "[B]oth compounds also show impressive effects in rodent models of age-associated diseases." (Kaeberlin, 2010, p. 96). This could be because they can modulate cell signaling close to a fasting state (e.g., resveratrol: Chatam et al., 2022; rapamycin: Blagosklonny, 2019). A sustainable alternation of exercise, IF or CR and the use of CRMs might lead to benefits similar to those of the continued reliance upon a single modality of NSPAN intervention. In this contribution, the potential benefits of EIC and CRMs will be dwelved upon, including the metabolic and signaling processes trough which these advantages are achieved.

## Neurodegenerative diseases and depression

The large majority of the papers mentioned in this review mention the devastating consequences of aging as one of the motors of research in the neurotrophic effects of EIC and CRMs, including NSPAN. A variety of essential restorative functions are decreased with age, which leads to brain aging (e.g., Zia et al., 2021). In addition, aging increases damage to DNA and tissues, including in neurons, stem cells and their niches, (Brunet et al., 2022; Maharajan et al., 2020). A bibliometric analysis of the literature on cognitive aging is provided by Othman et al. (2022). Some of the consequences of aging are due to epigenic changes in signaling pathways (e.g., Iside et al., 2020; Wang et al., 2022a). For instance, aging-related impaired AMPK signaling (see below) is related to atherosclerosis and cardiovascular diseases, increased risks of insulin resistance and diabetes, obesity, chronic inflammation, nonalcoholic fatty liver diseases and polycystic ovary syndrome, cancer and neurodegenerative diseases (Yadav et al., 2017). Neurodegenerative diseases share some ethiological processes, such as oxidative stress, post-translational modifications, metabolic changes, cell death mechanisms, disturbed immune response (Balusu et al., 2023; Fan et al., 2017). These factors have (epi)genetic causes and environmental and behavioral causes (Fan et al., 2017).

The most frequent neurodegenerative diseases are Alzheimer's disease and Parkinson's disease (McGregor & Nelson, 2019). Alzheimer's disease is characterized by an aggregation of amyloid beta peptide and accumulation of misfolded tau proteins, which can result from epigenetic dysregulations of cell signaling, including mitochondrial and metabolic disfunctions (Aman et al., 2021; Amorim et al., 2022; Balusu et al., 2023; Bloom, 2014; Guo et al., 2022). Examples of dysregulated processes linked to neurodegenerative impairments in Alzheimer's disease are endocytosis, proteastasis and phagocytosis in microglia, inflammation and metal ion homeostasis in astrocytes, as well as myelin loss and lipid deposits in oligodendrocytes (Balusu et al., 2023). Parkinsons' disease is characterized by the presence of Lewy bodies in different areas of the brain and neurodegenerative cell loss, dopaminergic neurons in particular (Balusu et al., 2023; McGregor & Nelson, 2019). Examples of dysregulated processes associated with neurodegeneration in Parkinsons' disease are: proteostasis, potassium ion transport and phagocytosis in the microglia, heat-shock response and metal ion homeostasis in astrocytes, as well as proteostasis and zinc homeostasis in oligodendrocytes (Balusu et al., 2023).

Major depression is the most important factor associated with disability among all causes (Craske et al., 2017). The search for biomarkers of major depression is challenging due to the heterogeneity of the description of the disorder (e.g., increased appetite/sleep, decreased appetite/sleep; decreased emotionality, increased nervousness; feeling slow, feeling restless); dysregulated cell signaling protein pathways in relation to inflammation and metabolism are candidate biomarkers for major depression (van Haeringen et al., 2022).

Progress stemming from a focus on brain circuits has been disappointing in research, prevention and treatment of neurodegenerative disorders and depression and some authors point out the relevance of examining the etiology and treatment of neurodegenerative diseases at the molecular and cellular level (Balusu et al., 2023). EIC and CRMs have the potential to maintain healthy cell signaling and restore dysregulated cell signaling (e.g., Bayliss et al., 2016; Erbaba et al., 2021; Heberden, 2016; Igwe et al., 2021; Kallies et al., 2019; Magliulo et al., 2022; Ntsapi & Loos, 2016; Overall et al., 2016; Radak et al., 2020; Ravula et al., 2021; Spasić et al., 2009; Szwed et al., 2021; van Praag et al., 2008; van Praag et al., 2014; Vints et al., 2022; Walsh et al., 2020). EIC and CRM are discussed next.

**Exercise**

The American College of Sports Medicine (ACSM, 2013) defines physical exercise as planned voluntary physical activity performed with the aim of maintaining or improving the fitness of the body. There are several types of exercise. Distinctions include: aerobic vs anaerobic exercise and strength vs endurance training (ACSM, 2013). In aerobic exercise, the body relies on aerobic metabolism (oxygen provided by pulmonary respiration) to generate the adenosine triphosphate (ATP) necessary for the activity, whereas in anaerobic exercise anaerobic glycolysis is relied upon for ATP production. Strength training aims to increase the force of the muscle. Endurance training aims to increase the potential duration of exertion (ACSM, 2013). While exercise can reduce oxidative stress and chronic inflammation (e.g., Beavers et al., 2010; Gomez-Cabrera et al., 2008), exercise is also a source of reactive oxygen species (ROS) and thereby can increase oxidative stress (Bloomer et al., 2005; He et al., 2016) and the antioxidative benefits of exercise depend upon its duration, type and intensity (Fisher-Wellman & Bloomer, 2009; Gomez-Cabrera et al., 2008). The importance of the redox balance is relation to exercise is discussed in Sutkowy et al. (2021).

Exercise can still provide benefits even when it is oxidative. These benefits include the activation of 5' adenosine monophosphate-activated protein kinase (AMPK, see below) (Richter & Hargreaves, 2013; Richter & Ruderman, 2009) and of sirtuins (SIRTs) (Radak et al., 2020). In addition to these, the indirect increase in the concentration of brain derived neurotrophic factor (BDNF) is also a potential consequence of exercise (Leckie et al., 2014). "BDNF mediates beneficial effects of energetic challenges such as vigorous exercise and fasting on cognition, mood, cardiovascular function and peripheral metabolism." (Marosi & Mattson, 2014, p.89). But these benefits can be balanced by the inflammation stemming from oxidative stress. In the recent years, the important role of exerkines – defined as "any humoral factors secreted into circulation by tissues in response to exercise" (Magliulo et al., 2022, p. 105) – has been highlighted in the explanation of the benefic effects of exercise (e.g., Chow et al., 2022; Magliulo et al., 2022; Reddy et al., 2022; Vints et al., 2022). Vints et al. (2022) discuss the implications of exercises (and exerkines) with regards to synaptic plasticity. A recent meta-analysis (Pillon et al., 2020) has focused on studies relying upon transcriptomic profiling to identify genes of which the expression is modified by exercise in the skeletal muscles. The mechanisms at play in

exercise–induced tissue regeneration including in the central nervous system (neurogenesis as well as neuron myelin and axon regeneration for instance) are described by Chen et al. (2022a). Cortes and De Miguel (2022) highlight the sex differences in CNS responses to exercise.

**Intermittent fasting**

The overconsumption of food is an issue in modern societies because of its constant availability. Discussing the detrimental effects of overconsumption for cognitive functioning and the damaging implications regarding neurodegenerative diseases, Mattson (2019) specifies that "[t]he underlying molecular mechanisms involve epigenetic modifications (…), that alter the expression of genes involved in neuroplasticity" (p. 204) and lead to "synaptic dysfunction, impaired neurogenesis, [and] neuronal degeneration" (p. 205). Nutritional practices that re–establish "adaptive cellular signalling," such as IF and CR, can mitigate the negative effects of food overconsumption on cognition by enhancing NSPAN (Mattson, 2019).

"Intermittent fasting (IF) is a regimen in which there are repeating cycles of ad libitum eating and fasting" (Erbaba et al., 2021, p. 2). Four main forms of IF are frequently studied (in addition to religious forms of fasting): alternate day fasting (important lowering of energy intake every two days), whole day fasting (one or two days of food abstinence a week) and time restricted feeding (energy intake limited to a time window of 8 to 12 hours, generally) and periodic fasting (were 5 days of ad libitum diet are followed by two days of fasting) (e.g., Bhoumik et al., 2023; Li et al., 2023; Seidler & Barrow, 2022; Tinsley & La Bounty, 2015). Similar benefits as those of CR (e.g., improved neurogenesis, Baik et al., 2020), which are discussed next, are observed for IF and fasting mimicking diets (Baik et al., 2020; Seidler & Barrow, 2022; Zhao et al., 2022), which consist of limiting the intake of calories and proteins for cycles of a few days to a week (see Seidler & Barrow, 2022). Chen et al. (2022a) provide a bibliometric analysis on intermittent fasting. Li et al., (2023) provide a balanced view of intermittent fasting (including potential disadvantages) and summaries of processes relating IF to circadian rhythms.

**Caloric restriction**

CR entails the limitation of energy intake from food – "A rough guideline would be 1800–2200 kcal/day for men and 1600–2000 calories for women" (Mattson, 2010, p. 8) – for a prolonged period of time (Mattson, 2010). The availability of nutrients (such as the ratio of adenosine monophosphate over ATP; Yadav et al., 2017) regulates several signaling pathways of relevance for NSPAN. Through various mechanisms, CR can delay the cognitive effects of aging (Gillespie et al., 2016). For instance, CR leads to the reduction of oxidative stress (Almendariz–Palacios et al., 2020; Yu et al., 2020a), to the proliferation of new neurons (Maharajan et al., 2020; Mattson & Arumugam, 2018), and to a better regulation of autophagy of neurons, i.e., the recycling of the material of deficient or damaged cells which is increased when energy is scarce (in the Appendix, see the summary of Ntsapi & Loos, 2016). Other examples include: the increase of anti–inflammatory hormones, the decrease of chronic inflammation, and the modulation of cell survival (e.g., through apoptosis) (Testa et al., 2014). Some of the signaling pathways involved in these processes and others are described in the next section.

**Caloric restriction mimetics**

CRMs share some of the benefits of EIC (Testa et al., 2014; Xue et al., 2021). For instance, Bonkowski & Sinclair (2016; see Appendix) discuss the increasing use of nicotinamide adenine dinucleotide precursors and activators of SIRTs in clinical trials in the context of (neurodegenerative) disease prevention and treatment and the promotion of longevity. Several reviews summarized in the Appendix discuss the role of polyphenols in the modulation of synaptic plasticity, increased neuronal proliferation, and the reduction of oxidative stress and of neurotoxicity in particular (e.g., Moosavi et al., 2016; Rendeiro et al., 2015; Zhang et al., 2021). It should be noted that Miller et al.

(2011) found that resveratrol and simvastatin did not provide longevity benefits in rodents, whereas rapamycin did. These three CRMs thus impacted longevity differently in Miller et al. (2011) (but see Madeo et al., 2019). Sharman et al., (2019) note a low success rate of phytocomponents (among which nicotinamide adenine dinucleide boosters and sirtuins upstream effectors) in terms of symptom reduction in human trials of Alzheimer's patients. Importantly, the safety of nicotinamide mononucleotide, a nicontinamide adenine dinucleotide precursor, needs to be further investigated (Nadeeshani et al., 2022).

**Neurogenesis**

Adult neurogenesis refers to the formation of new neurons in the adult central nervous system (Overall et al., 2016). A description of the different stages of neurogenesis, starting with the stem cell and leading to the mature neuron, is provided by Overall et al. (2016). New neurons are hyperexcitable and less receptive to inhibitory neurotransmitters (e.g., gamma–Aminobutyric acid – GABA), which confers on them more influence than older neurons in a neural circuit (Fares et al., 2019). Adult neurogenesis has been documented in different mammals (Lucassen et al., 2020). Studies on the matter overwhelmingly conclude that neurogenesis occurs in the adult human hippocampus (see Lucassen et al., 2020; Kempermann et al., 2018). The hippocampus fulfills important functions with respect to spatial and episodic memory and is involved in emotional processes (Anacker & Hen, 2017). Neurogenesis in this region – accounting to 700 new neurons every day (2% of the total; Spalding et al., 2013) – might be required for flexible representations and behavior, i.e., in dealing with new situations adaptively (Tuncdemir et al., 2019). Reduced hippocampal neurogenesis is associated with diminished ability to form new memories and the forgetting of old memories (Frankland et al., 2013), whereas increasing hippocampal neurogenesis allows for more flexibility in strategy choice, and can compensate preceding deficiencies in the aging brain (Berugo–Vega et al., 2020).

Adult neurogenesis can be induced or enhanced through EIC and CRMs, which slows down cognitive decline and can also enhance cognition (e.g., Chow et al., 2022; Mattson & Arumugam, 2018). Overall et al. (2016) describe the different mechanisms through which neurogenesis is promoted by physical activity. Maharajan et al. (2020) and Negredo et al. (2020) discuss the potential of CR in maintaining the health of stem cells and stem cell niche regeneration. Fares et al. (2019) provide a historical perspective on the study of neurogenesis.

**Neuroprotection**

EIC and CRMs can promote neuroprotection and thereby delay cognitive aging (e.g., Burtcsher et al., 2022; Mattson & Arumugam, 2018; Reddy et al., 2022). This is achieved through several mechanisms. Two are addressed in this paragraph, while others are briefly mentioned in the pages that follow. The first of these mechanisms is the protection of the neuron and of the micro–environment of neurons from oxidative stress (e.g., protection of membrane and DNA integrity; Li et al., 2022; Lüscher et al., 2021). The second related mechanism is the regulation of cell survival (e.g., Ying, 2006): In ideal conditions, the survival of healthy cells is promoted, while the survival of senescent and otherwise deficient cells can be impeded through different mechanisms (e.g., autophagy and apoptosis) (e.g., Bortner & Cidlowsky, 2020; Laberge et al., 2013), with neuroprotective effects thereby achieved in the brain (Bhaduri et al., 2023). At the cellular level, different molecules and signaling pathways play a role in determining the fate of the cells, including neurons (e.g., Bathina & Das, 2015; Fayad et al., 2005; Huang & Reichardt, 2001; Jaworski et al., 2019; Krieglstein et al., 2002; Li et al., 2022; Lüscher et al., 2021; Nordvall et al., 2022; Rai et al., 2019; Ying, 2006; Yamamoto et al., 2007; Zhao et al., 2019). EIC and CRMs can regulate such processes (e.g., Mattson & Arumugam, 2018).

**Synaptic plasticity**

Synaptic plasticity is also an important neurotrophic effect of EIC and CRMs (e.g., Reddy et al., 2022; Rendeiro et al., 2015; Seidler & Barrow, 2022). The ability of the brain to reorganize following internal and external events is key to learning and adaptive conduct; one such mechanism is synaptic plasticity (Bliss et al., 2018; Magee & Grienberger, 2020; McAllister et al., 1999). Neurotrophins (in particular BDNF, see below in the current section) play an important role in synaptic plasticity as well as synaptic transmission (Gómez-Palacio-Schjetnan & Escobar, 2013; McAllister et al., 1999). Magee and Grienberger (2020) discuss several forms of synaptic plasticity, including the form most commonly studied: Hebbian plasticity (the strength of the synapse depends on how often the presynaptic neuron fires toward the postsynaptic neuron). Abraham (2008) discusses meta-plasticity, which they refer to as the tuning of synapses and neuron networks for plasticity. "Metaplasticity entails a change in the physiological or biochemical state of neurons or synapses that alters their ability to generate synaptic plasticity" (Abraham, 2008, p. 387): Alternations between phases of high stimulation and phases of low stimulation are paradigmatic of the promotion of meta-plasticity, and physical exercise can induce high stimulation.

## B) Energy Metabolism, Oxidative Stress, Neuroprotection, Synaptic plasticity and Adult Neurogenesis: Biochemical Processes

In this section, several of the processes that are responsible for the effects of EIC and CRMs on NSPAN are discussed. Further, other neurotrophic and protective effects of signaling pathways that are elicited by EIC and CRMs are presented for comprehensiveness.

### Energy metabolism

Liver glycogen storage provides the body with continued access to 'fuel' as long as the supply lasts through the release of glucose to the bloodstream (McBride & Hardie, 2009). Glucose transporters are proteins that mediate the entry of glucose across cell membranes and the blood brain barrier (BBB) (e.g., Simpson et al., 1999; Thorens & Mueckler, 2010). Cellular respiration results in the production of ATP in cell mitochondria (Korla & Mitra, 2014). This process also releases $H_2O$ and $CO_2$. Cellular respiration starts with glycolysis (the transformation of glucose to pyruvate, then to acetyl-coenzyme-A; Acetyl-CoA), of which the availability of nicotinamide adenine dinucleotide is an important requirement. It continues with the Krebs cycle, and ends with oxidative phosphorylation – itself involving the electron transport chain (ETC) and chemiosmosis (Zhao et al., 2019).

When energy is necessary in the cell, notably through hydrolysis (the binding of $H_2O$ with ATP), ATP is converted to: adenosine diphosphate (ADP) + one phosphate that can be consumed as energy (Yadav et al., 2017). The same can occur for ADP, leading to adenosine monophosphate (AMP) + one phosphate. The ratios ADP:ATP and AMP:ATP are thus indicators of the energy requirements in the cell (Yadav et al., 2017).

Hydrolysis of triglycerides occurs when ADP:ATP or AMP:ATP ratios are high and results in the synthesis of fatty acids, which are then converted to ketone bodies through oxidation to fuel the cells (g-to-k metabolic switching; Martínez-Reyes & Chandel, 2020; Mattson et al., 2018; Newman & Verdin, 2017).

Energy expenditure (e.g., exercise), increases ROS in the organism (He et al., 2016). A major contributor of ROS is the slippage of electrons through the ETC in the mitochondria (Mazat, Devin & Ransac, 2020; Shofield & Schafer, 2021; Zhao et al., 2019). Proton slippage is partially down-regulated by uncoupling proteins (UCPs; Demine et al., 2019; Haas & Barnstable, 2021). Holmström & Finkel (2014) discuss other sources of ROS. The processes of ROS generation and the role of antioxidants and ROS inhibitors are reviewed by Brieger et al. (2012).

Oxidative stress ensues from an elevated production of ROS coupled with insufficient availability of antioxidants (ROS imbalance) (Aon et al., 2010; Shofield & Schafer, 2021). "As signaling molecules, ROS play an important role in cell proliferation, hypoxia adaptation and cell fate determination, but excessive ROS can cause irreversible cell damage and even cell death" (Zhao et al., 2019, p. 3). Oxidative stress more generally can also lead to such negative consequences (Murphy, 2009; Di Meo et al., 2016; Shofield & Schafer, 2021). Zeliger (2013) discusses diseases linked with oxidative stress, their comorbidity and prevention, as well as the Oxidative Stress Index questionnaire, which permits to assess the potential future onset of non-communitive chronic diseases and a list of environmental sources of ROS.

A moderate ROS is ideal for the endocrine, immune and cognitive functions (Brieger et al., 2012) and thus both low and high ROS are related to metabolic disfunctions and diseases (Brieger et al., 2012). Importantly, an elevated ROS is associated with a range of neurodegenerative diseases (Singh et al., 2019). Dunn et al (2015) present the anti-microbial function of ROS. Holmström & Finkel (2014) discuss redox signaling. Chun et al. (2010) discuss the antioxidant intake of Americans from several sources.

Both high AMP:ATP ratio and high ROS are sensed by the 5' adenosine monophosphate-activated protein kinase (AMPK) and trigger its signaling (e.g., Hinchy et al., 2018; Rabinovitch et al., 2017). Different protective mechanisms are thereby indirectly activated by increased ROS and involve SIRTs, peroxisome proliferator-activated receptor-γ coactivator (PGC)-1α, mammalian target of rapamycin (mTOR), Unc-51 Like autophagy activating Kinase 1 (ULK1), transcription factors forkhead box O (FOXO) and p53. Promyelocytic leukemia (PML) is a sensor of ROS; when ROS is sensed, it activates p53 (Niwa-Kawakita et al., 2017). P53 is an inhibitor of cell growth and propagation and is involved in tumor suppression (Aubrey et al., 2018; Budanov & Karin, 2009). It is involved in neuroprotection by regulating neuronal apoptosis (Culmsee & Mattson, 2005). Maor-Nof et al. (2021) note that p53 is linked with neurodegeneration.

## Nucleotides, Metabolites and Neurotrophic Factors

### *Nicotinamide adenine dinucleotide*

NAD+ is involved in DNA repair and mitophagy (the process of recycling of non-essential or damaged mitochondria; Schofield & Schafer, 2021) as it is required for SIRT1 activation (Chen et al., 2020; Fang & Bohr, 2017). NAD+ participates in glycolysis and is reduced to NADH in the process (Navas & Carnero, 2020; Ying, 2006).  As mentioned above, a high level of ROS is an important threat to mitochondrial and cellular integrity. Increased NAD+ can up-regulate immunity, inflammation and DNA repair through the activation of NAD+ dependent enzymes such as SIRTs and poly (ADP-ribose) polymerases (PARPs) (Navas & Carnero, 2020; Xie et al., 2020; Ying, 2006). Although it has been reported that NAD+ acts as a proinflammatory cytokine (e.g., Navas & Carnero, 2020; Xie et al., 2020), it has also been shown that NAD+ and its precursors can regulate chronic inflammation and reduce ROS, and, thereby, inflammation (Covarrubias et al., 2021; Xie et al., 2020). NAD+ also modulates the effect of different enzymes, facilitates energy metabolism and helps in the synchronization of the circadian clock (Xie et al., 2020). NAD+ and NADH also play important roles in cell survival and death (Ying, 2006), including through their role in ADP-ribosylation (Li et al., 2022; Lüscher et al., 2021). Aging is related to NAD+ deficiency and such deficiency is related to accelerated aging (e.g., increased mitochondrial disfunctions; Xie et al., 2020). This constitutes a vicious circle – which can be thwarted for instance by supplementation of NAD+ precursors (Bonkowski & Sinclair, 2016) and the up-regulation of AMPK signaling through exercise (Richter & Hargreaves, 2013) and calorie restriction (Mattson & Arumugam, 2018). Consequences include disturbed circadian rhythms, increased inflammation, poor immunity, hampered

DNA repair, and increased risk of cancer, which is aggravated by high levels of ROS (Navas & Carnero, 2020; Xie et al., 2020).

Three pathways are at play in the synthesis of NAD+: the de novo, Preiss–Handler and Salvage pathways (Lautrup et al., 2019; Xie et al., 2020; Ying, 2006). Synthesis of NAD+ through the de novo pathway has neuroprotective and neurotoxic effects (Lautrup et al., 2019). It "starts with the catabolism of the amino acid tryptophan that is converted via two steps to the intermediate kynurenine, which can generate NAD+, kynurenic acid, or xanthurenic acid." (Lautrup et al., 2019, p. 630). NAD+ is synthesized from nicotinic acid (NA) through the Preiss–Handler pathway and from nicotinamide riboside (NR) and the reuse of nicotinamide (NAM) through the Salvage pathway (Lautrup et al., 2019; Xie et al., 2020; Ying, 2006). Both the Preiss–Handler and Salvage pathways rely upon the intermediary synthesis of nicotinic acid mononucleotide (NAMN, Lautrup et al., 2019; Xie et al., 2020; Ying, 2006). It is worth noting that "ATP may be consumed to replete NAD+ when NAD+ is selectively depleted by enzymes such as PARP–1; vice versa, NAD+ may also be consumed to regenerate ATP under the pathological conditions when ATP is selectively depleted." (Ying, 2006). The therapeutic potential of nicotinamide riboside supplementation is examined by Mehmel et al. (2020).

## Ketone Bodies

Fatty acids are synthesized (through hydrolysis of triglycerides) when glucose is lacking, for instance following EIC, and then converted into the three ketone bodies: β–hydroxy–butyrate (BHB), acetone (Ace) and acetoacetic acid (also referred to as acetoacetate, AcAc) (Newman & Verdin, 2017). Ketone bodies are then distributed to the cells of the organs, including the brain, through their release in blood circulation. Ketone bodies can be used as alternate fuel sources to glucose: In the cell, they are catabolized into Acetyl–CoA (Newman & Verdin, 2017). Acetyl–CoA is then converted to ATP as a result of the Krebs cycle (for details, see Martínez–Reyes & Chandel, 2020).

Ketone bodies also have regulatory roles, such as the response to oxidative stress, cell excitability, the regulation of protein expression, and signaling more generally in and out of the brain (see García–Rodríguez & Giménez–Cassina, 2021; Koppel & Swerdlow, 2018). There is evidence that increases in ketone bodies, BHB in particular, result in neuroprotection (e.g., Kolb et al., 2021). BHB can be transported across the BBB by monocarboxylic acid transporters (MCT). The amount of MCT regulates BHB in the brain, where it is responsible for a variety of actions (Newman & Verdin, 2017). Some of these are briefly mentioned: An important role of BHB is the upregulation, through activation of NF–κB, of the hippocampal expression of BDNF in case of either scarce or sufficient glucose supply (Hu et al., 2018; Mattson & Arumugam, 2018). In other words, BHB increases BDNF even in the absence of g–to–k metabolic switching. BHB affects sympathetic tone through inhibition of the free fatty acid receptor 3 (Newman & Verdin, 2017). It up–regulates gene expression and suppresses oxidative stress by inhibiting class I histone deacetylases (HDAC; Newman & Verdin, 2017; Shimazu et al., 2013). It appears BHB is capable of stimulating PGC–1α (mitochondrial biogenesis and mitophagy) and FOXO1 (metabolic regulation, reduced sympathetic tone, apoptosis; Santo & Paik, 2018) without AMPK modulation through the Phosphoinositide 3–kinase – Protein kinase B (PI3K/Akt) pathway (Kim et al., 2019). Through the activation of macrophages, BHB and nicotinic acid (NA), activate the hydroxy–carboxylic acid receptor 2 (HCAR2) which induce neuroprotection through the reduction of neuroinflammation and also regulates lipid metabolism (Newman & Verdin, 2017; Rahman et al., 2014). BHB is also involved in synaptic plasticity (Mattson & Arumugam, 2018) and GABAergic regulation (Newman & Verdin, 2017) and plays a major role in cognition and health in general (Cavaleri & Bashar, 2018).

## Brain Derived Neurotrophic Factor (BDNF)

BDNF, a member of the neurotrophins family, is a main protagonist in NSPAN that can be increased through exercise (Bathina & Das, 2015; Kallies et al., 2019 ; Zhao et al., 2008). Neurotrophins are a family of proteins that regulate neural function, growth and survival, and neuroplasticity through various signaling pathways (Huang & Reichardt, 2001; Bathina & Das, 2015). There are four neurotrophins: nerve growth factor (NGF), neurotrophin-3 (NT-3), neurotrophin-4 (NT-4) and BDNF. The role played by BDNF in regulating neurophysiological processes, modulating synaptic interactions, neuroplasticity and neuroprotection has been extensively studied (Kowiański et al., 2018). Such functions are mediated by several downstream effectors, including the Trk family of tyrosine kinase family of receptors (Foltran & Diaz, 2017). Activation of the tyrosine receptor kinase B (TrkB) kinase promotes signaling along two pathways: PI3K/Akt and Raf/MEK/ERK (Foltran & Diaz, 2017). A few examples of the more specific roles of BDNF are now provided. Research has found that BDNF increases axonal arborization through tyrosine receptor kinase B (TrkB) signaling, a process which is regulated by the phosphitylation of the mitogen-activated protein kinases (MAPK; Cheng et al., 2011; Jeanneteau et al., 2010). TrkB is further involved in neural plasticity (Johnstone & Mobley, 2020). BDNF also increases the activation of neuronal stathmins (Cardinaux et al., 1997). Stathmins induce plasticity, the generation of new neurons and their development (Chauvin & Sobel, 2015). It is notable that BDNF also up-regulates mTORC1, which can be detrimental; but this action is inhibited by AMPK and its effect on SIRT signaling through the up-regulation of NAD+ metabolism (Ishizuka et al., 2013; Sadria & Layton, 2021). Reduced levels of BDNF are associated with neurodegenerative diseases (for reviews, see: Bathina & Das, 2015; Autry & Monteggia, 2012 – who mention its relation with neurotransmitter serotonin). In the body and in the brain, BDNF is also involved in energy homeostasis and regulation of the parasympathetic tone (Marosi & Mattson, 2014).

Interventions targeting neurotrophins, including BDNF are discussed by Nordvall et al. (2022): agonists of TrkB and BDNF are explored in such interventions, e.g.: steroid derivatives, "gambogic amide, asiaticoside, and sarcodonin" (p. 3) as well as 7,8-dihydroxy flavone (a polyphenol; see Emili et al., 2022). Several of which did not show robust effectiveness, which questions their validity (Nordvall et al., 2022).

**Protein and Lipid Kinases Involved in NSPAN**

Kinases are enzymes which add a phosphor group to their target when activated (phosphorylation, Taylor & Kornev, 2011). In this sub-section, several protein and lipid kinases involved in NSPAN are briefly presented. The main actions of these effectors are summarized in Table 1.

*Phosphoinositide 3 (PI3K)*

IF and CR are involved in the improvement of mitochondrial function through the regulation of PI3K and its effectors (Lu et al., 2019; Zhao et al., 2022). PI3K dysregulation is implicated in neurodegenerative diseases such as AD (Kumar & Bansal, 2022), which has been associated with increased chronic microglia activation following such dysregulation (Chu et al., 2021): "considerable evidence suggests that chronically activated microglia continually secrete neurotoxic molecules to sustain neuroinflammation" (p. 3). Further, PI3K is involved in the regulation of autophagy, cellular growth and survival and synaptogenesis, notably through its direct or indirect influence on Akt, mTOR (PI3K/Akt/mTOR pathway, Ersahin et al., 2015) and ERK (Carracedo & Pandolfi, 2008; Chen et al., 2013; Cuesto et al., 2011; Rai et al., 2019; Sadria & Layton, 2021). Key activators of PI3K include insulin/insulin growth factor 1 (IGF-1), the tyrosine receptor kinase B (TrkB), and BHB (Foltran & Diaz, 2017; Kim et al., 2019; Sadria & Layton, 2021). PI3K is an upstream regulator of Ras (Bathina & Das, 2015), which itself regulates Raf, the entry point of the Raf/MEK/ERK pathway (Peyssonnaux & Eychène, 2001).

PI3K is up-regulated by insulin receptor activation, IGF-1 and TrkB activity in particular (Foltran & Diaz, 2017; Sadria & Layton, 2021). Fakhri et al. (2021) examine the use of natural molecules to reduce the activity of the PI3K/Akt/mTOR signaling pathway. A history of research on PI3K is provided by Vanhaesebroeck et al., (2012) and a bibliometric analysis of the literature on PI3K by Ho & Hartley (2020).

### Protein kinase B (Akt)

Akt dysregulation is implicated in neurodegenerative diseases such as AD (Kumar & Bansal, 2022). Akt is involved in neuroprotection in several ways as: up or down-regulation of the functions of other proteins responsible for apoptosis (programmed death of deficient cells), and survival and proliferation of cells – including autophagy, notably neurons and the regulation of neuronal toxicity (Chen et al., 2013; Fayad et al., 2005; Rai et al., 2019). Examples include the down-regulation of AMPK – through a decrease in the AMP:ATP ratio and the up-regulation of mTOR (Hahn-Windgassen et al., 2005). Fayad et al. (2005) mention that other pathways may also regulate pathways modulated by Akt. Akt also phosphorylates FOXO and the tuberous sclerosis complex (TSC1 and TSC2, important regulators of mTOR; Huang et al., 2008; Sadria & Layton, 2021) and regulates the anti-apoptotic and pro-survival extracellular signal-regulated kinases (ERK) (Rai et al., 2019). Vadlakonda et al. (2013) discuss the interplay ('interactions and feedback loops') between Akt and mTOR and the mediating role of AMPK and FOXO in particular.

The Insulin/insulin growth factor 1 (IGF-1) pathway activates phosphoinositide-dependent kinase-1 which activates Akt (Sadria & Layton, 2021). Akt activation can also be promoted through PI3K and mTORC2 (Huang et al., 2008; Sadria & Layton, 2021; Esrahin et al., 2015; Vadlakonda et al., 2013).

### Mammalian target of rapamycin (mTOR)

mTOR is related to two complexes: mTORC1 and mTORC2. The TSC1-TSC2 complex down-regulates mTORC1 and up-regulate mTORC2 (Huang et al., 2008). Neurodegenerative diseases are a potential consequence of chronic activation of mTORC1 (Sardia & Layton, 2021). mTOR is inhibited by CR and IF due to their effect on AMPK (Zhao et al., 2022). While exercise has the potential to increase mTOR activation due to the activation of Akt, "far from increasing mTORC1 activity, during running, the high amounts of metabolic stress induced by endurance exercise can decrease mTORC1 activity. This decrease in mTORC1 activity following running may be the result of the activation of the 5'-adenosine monophosphate activated protein kinase (AMPK)" (Watson & Baar, 2014, p. 134). mTOR is both a downstream and upstream effector of Akt. mTOR is involved in sensing "nutrient levels, the presence of growth factors, and the cellular energy status and mediates several catabolic and anabolic processes to maintain metabolism and cell growth" (Ersahin et al., 2015, p. 1947). Specifically, mTOR participates in autophagy (the recycling of cell materials and potentially cell death), the synthesis of proteins and nucleotides as well as glucose, mitochondrial and lipids metabolism (Chen et al., 2013; Liu & Sabatini, 2020; Szwed et al., 2021). When mTORC1 is activated chronically (and potentially in case of severe and prolonged acute activation), the risk of cancer is increased (Sadria & Layton, 2021).

mTORC1 can be directly inhibited by rapamycin (Szwed et al., 2021). It is less active when amino acids are low (Szwed et al., 2021). In particular, its activation results from the binding of arginine with Cytosolic arginine sensor for mTORC1 subunit 1 (CASTOR1) (Budanov & Karin, 2009; Chantranupong et al., 2016; Szwed et al., 2021) and leucine with SESN 2 (Budanov & Karin, 2009; Chen et al., 2022b; Szwed et al., 2021). The phosphorylation of ULK1 by AMPK down-regulates mTORC1, while the phosphatidylinositol 3-kinase/Akt pathway up-regulates mTORC1 (Szwed et al., 2021; Sadria & Layton, 2021). Unlike mTORC1, mTORC2 can not be inhibited by rapamycin directly (Szwed et al., 2021). PI3K activates mTORC2 (Sadria & Layton, 2021). mTORC2 promotes Akt activation through different means (Huang et al., 2008; Sadria & Layton, 2021; Esrahin et al., 2015).

Caron et al. (2010) provide a summary of the upstream and downstream effectors of mTOR. A bibliometric analysis of the literature on mTOR is provided by Li et al (2022).

*5' adenosine monophosphate–activated protein kinase (AMPK)*

AMPK is a considered target in anti-aging approaches aiming at neuroprotection, including the prevention and the alleviation of neurodegenerative diseases (e.g., Bayliss et al., 2016). Primary functions of AMPK are the sensing of cellular energy (e.g., high AMP:ATP ratio, low ATP), and the metabolic regulation that ensues; as well as the regulation of mitochondrial ROS (Rabinovitch et al., 2017; Zhao et al., 2019). Other important functions include a) the subsequent increase or decrease in the oxidized form of nicotinamide adenine dinucleotide (NAD+) and sirtuin–1 (SIRT1, Cantó et al., 2009), b) the down or up-regulation of pathways that consume or produce ATP (Cantó et al., 2009) and c) mitochondrial biogenesis (Cantó et al., 2009), neuron cell fate (Spasić et al., 2009). The role of AMPK in cellular metabolism (e.g., fatty acid synthesis, protein synthesis, glycolysis, glucose transport, endothelial nitride oxide synthase) is discussed by Yadav et al. (2017). Importantly, with the result of restoring energy balance in the cell, AMPK upregulates PGC–1α, phosphorizes ULK1 and tuberous sclerosis complex (TSC) 2 and thereby suppresses mTOR signaling (Rabinovitch et al., 2017; Sadria & Layton, 2021), an action which requires the activation of protein kinase B (Akt, Alessi et al., 2006); and SIRT directly (Yadav et al., 2017) and indirectly through increasing NAD+ (Tang et al., 2016). AMPK serves several other functions, including, among others, the regulation of autophagy (Karabiyik et al., 2021; Penugurti et al., 2022).

In addition to regulation related to energy sensing, AMPK activity is inhibited by ULK1 (Sadria & Layton, 2021) and is activated by cellular stress, leptin/adiponectin, fasting and exercise (Chen et al., 2022a; Hinchy et al., 2018; Richter & Hargreaves, 2013; Richter & Hargreaves, 2013; Yadav et al., 2017). P53 activates AMPK through Sestrin (SESN) 1 and SESN 2 (Budanov & Karin, 2009). SESNs are stress sensing proteins that participate in cellular homeostasis through the regulation of AMPK and mTOR and play an essential role in disease prevention (Chen et al., 2022b). p53 is activated as well by the tumour suppressor serine–threonine kinase liver kinase B1 (LKB1, Alessi et al., 2006; Shackelford & Shaw, 2009), which is also required for AMPK down–regulation of mTOR signaling (Shaw et al., 2004). A bibliometric analysis of the literature on AMPK is provided by Lyu et al (2022).

*Gylcogen Synthase Kinase–3β (GSK3β)*

The neurotrophic actions of GSK3β are mainy related to its role in the regulation of synaptic plasticity, synaptogenesis and neuronal proliferation and migration (Cuesto et al., 2015; Jaworski et al., 2019). GSK3β is also involved in cell survival and proliferation, metabolism and neuronal function (Jaworski et al., 2019). GSK3β plays an important role in the regulation of circadian rhythms through the activation and inactivation of nuclear factor erythroid 2–related factor 2 (Nrf2), which is  also activated by oxidative stress (Shilovsky et al., 2021). Nrf2 up–regulates stress adaptation through inhibition of inflammatory processes and ROS reduction (Shilovsky et al., 2021). Dysregulation of GSK3β has implications for neurodegenerative diseases and major depression (Jaworksi et al., 2019). GSK3β is down–regulated by phosphorylation (e.g., by Akt) and is activated by dephosphorization (e.g., by serine/threonine protein phosphatase 1) (Jaworski et al., 2019). IGF–1 up–regulates PI3K and Akt and down–regulates GSK3β (Duarte et al., 2008). Mutations along several pathways (PI3K, hedgehog, Notch, Wnt) are related to alterations in GSK3β activity (McCubrey et al., 2016). GSK3β phosphorylates β–catenin (Hermida et al., 2017), which is essential in the canonical actions of Wnt (Arredondo et al., 2020).

*Unc–51 like autophagy activating kinase (ULK)*

The ULK complex is composed of ULK1 and ULK2. ULK1 is involved in autophagy (recycling of cell materials) in conditions of cellular stress, which can affect the functioning of the cells and their survival or death (Nie et al., 2016). The ULK complex is also involved in mitophagy, which is the quality control process of the mitochondria with functions such as the "degradation of damaged and superfluous mitochondria, prevent[ion of] detrimental effects [of ROS and mitochondrial disfunction] and reinstat[ion of] cellular homeostasis in response to stress" (Chen et al., 2020, p. 1). Disfunctions in autophagy and mitophagy are importantly linked to neurodegenerative disorders (e.g., Chen et al., 2020). ULK is up-regulated by AMPK and inhibited by mTOR (Sadria & Layton, 2021; Yu et al., 2021).

### *Mitogen-activated protein kinases (MAPK) and signal cascades*

MAPK signal cascades play an essential role in the expression of some genes, immunity, cell differentiation and proliferation and are also involved in regulating neuronal inflammation and axon arborization (Ersahin et al., 2015; Peyssonnaux & Eychène, 2001; Wancket et al., 2012). Among those, an important signal cascade is the Raf/MEK/ERK pathway (Peyssonnaux & Eychène, 2001). Raf/MEK/ERK is also involved in cellular apoptosis and thereby regulates cell survival (Peyssonnaux & Eychène, 2001). This signaling pathway is complementary to the PI3K/Akt/mTOR pathway (Ersahin et al., 2015).

However, "MAPKs are implicated in the pathogenesis of many conditions" (Wancket et al., 2012, p. 244). MAPK phosphatases (MKPs) down-regulate MAPKs and MAPKs increase the activity of MKPs (Wancket et al., 2012) this feedback loop allows for the beneficial functions of MAPKs to be achieved, while keeping their activity under control. Ras, and Raf indirectly, are up-regulated by PI3K leading to the activation of some MAPKs (e.g., ERK; Peyssonnaux & Eychène, 2001). "On the one hand, hypoglycemia [e.g., which can be achieved through EIC] decreases anabolic hormones (e.g. insulin, GH, and IGF1), as well as sex and thyroid hormones, increases the expression of the catabolic cortisol, and subsequently inhibits the MAPK pathway (i.e. RAS/RAF/MEK/ERK) and the PI3K/Akt/mTOR pathway. On the other hand, the inhibition of ERK avoids mTOR activation and subsequently induces autophagy activity, which contributes to the suppression of inflammation by downregulation of both IFN [i.e., interferons] and proinflammatory cytokine responses." (Kökten et al., 2021, p. 1561).

## Other Signaling Proteins Involved in NSPAN

Other signaling proteins which play an important role in NSPAN are presented here. Their main actions are summarized in Table 1.

### *Peroxisome proliferator-activated receptor-γ coactivator (PGC-1α)*

Exercise, calorie restriction and intermittent fasting, as well as exposure to cold temperature, up-regulates PGC-1α (Anderson & Prolla, 2009; Chen et al., 2022a; Vernier & Giguère, 2021). PGC-1α plays a major role in a reducing oxidative stress and inflammation and in regulating mitochondrial biogenesis and mitophagy (Rius-Pérez et al., 2020; Tang et al., 2016). It is also involved in fatty acid oxidation, glucogenesis and energy metabolism (regulation of the Krebs cycle and of ATP production) (Rius-Pérez et al., 2020). "Dysregulation of PGC-1α alters redox homeostasis in cells and exacerbates inflammatory response, which is commonly accompanied by metabolic disturbances." (Rius-Pérez et al., 2020, p. 1). Dysregulation of PGC-1α accrues with aging, and anti-aging intervention have been documented to target PGC-1α as to limit the impact of mitochondrial decline including on the nervous system (Anderson & Prolla, 2009; Vernier & Giguère, 2021).

PGC-1α is mainly regulated by AMPK (positively), Akt (negatively) and glycogen synthase kinase 3β (GSK3β; negatively) (Chen et al., 2022a; Rius-Pérez et al., 2020) as well as indirectly by SIRT (Cantó et al., 2013). BHB can up-regulate PGC-1α (Kim et al., 2019).

## Nuclear factor κB (NF-κB)

Moderate exercise might inhibit NF-κB, while resistance exercise activates it (Liu et al., 2018; Vella et al., 2012). NF-κB is an important regulator of the cell cycle

(thereby affecting cell fate), immunity and inflammation (Liu et al., 2017; Mattson & Arumugam, 2018). Such functions of NF–κB are relevant for neurodegeneration (Yu et al., 2020b). In particular, NF–κB chronic activation has been implicated in the pejoration of symptoms of Alzheimer's disease and neuroinflammation – a risk factor for neurodegenerative diseases and depression (Capece et al., 2022; Dolatschahi et al., 2021); and NF–κB is involved in contradictory functions such as cell survival and neurodegeneration (Mincheva–Tasheva et al., 2013): "the same stimulus produces opposite responses by activating NF–κB in different cell types" (p. 187). Some of the upstream effectors of NF–κB are: lactate, succinate, BHB, (un)saturated fatty acids) (a more complete list is provided by Capece et al., 2022). NF–κB is also indirectly activated by ROS (Liu et al., 2017).

## Sirtuins (SIRT)

SIRTs are a class of histone deacetylases (Carafa et al., 2016). "The seven members of this family of enzymes are considered potential targets for the treatment of human pathologies including neurodegenerative diseases, cardiovascular diseases, and cancer." Carafa et al., 2016, p. 1). The availability of NAD+ is required for the activation of all 7 SIRTs (Cantó et al., 2009; Tang et al., 2016; Elkhwanky & Hakkola, 2018; Fagerli et al., 2022). Such requirement fulfills energy sensing functions: Thereby, SIRTs can optimally respond to cellular energy requirements (Elkhwanky & Hakkola, 2018). SIRT1, SIRT6 and SIRT7 are present in the cell nucleus, whereas SIRT2 is present in cytosol and SIRT3, SIRT4, and SIRT5 are present in the mitochondria (Elkhwanky & Hakkola, 2018; Fargeli et al., 2022). Morris (2013; focus on aging) and Yamamoto et al. (2007) addresses in detail the functions of SIRTs, notably: gluconeogenesis, lipid metabolism, mitochondrial respiration, cellular proliferation and differentiation, cell cycle control, gene expression, apoptosis, inflammation, insulin secretion, neuroprotection and axonal protection, DNA repair. Elkhwanky & Hakkola (2018) discuss the important role of extranuclear SIRTs. SIRT1, of which the activity has been in particular associated with increased longevity (for details, see Tang et al., 2016) has been the focus of most research on SIRTs. In what follows, the activity of SIRTs is discussed in very general terms, as the signaling of SIRTs is especially complex and would require much space (but see Cantó et al., 2009, 2013; Tang et al., 2016; Elkhwanky & Hakkola, 2018; Fagerli et al., 2022). Specific mentions of SIRT signaling are found in the other sections.

Increased activity of SIRTs (e.g., by AMPK, or indirectly by exercise, CR or CRMs; e.g., Chen et al., 2022a) is related to energy homeostasis (e.g., regulation of the Krebs cycle and oxidative phosphorylation, triggering of glucose uptake, fatty acid oxidation and ketogenesis), mitochondrial homeostasis and integrity (notably: reduction of ROS and DNA repair through activation of PARPs, see below), and NSPAN (regulation of synaptic plasticity; mitochondrial biogenesis in neurons importantly; reduction of ROS) (Elkhwanky & Hakkola, 2018; Fagerli et al., 2022). Some of the functions in which SIRTs are indirectly involved in rely upon ADP–ribosylation signaling (see below) in which PARPs play a major role (Hottiger, 2015). A higher risk of neurodegenerative diseases ensues the decreased activity of SIRTs, such as mitochondrial and cardio–vascular diseases, and neurodegenerative diseases (Elkhwanky & Hakkola, 2018; He et al., 2022; Jiao & Gong, 2020). Importantly, SIRTs can activate LKB1, which in turn can activate AMPK and down–regulate mTOR (Sadria & Layton, 2021; Shaw et al., 2004; Woods et al., 2003) and regulate PGC–1α (Cantó et al., 2013). The modulation of SIRTs, for instance using resveratrol among other SIRT activators, is a promising prevention and treatment target for neurodegenerative disorders and cancer (Cao et al., 2018; Carafa et al., 2016). Leite et al. (2021) discuss the relevance of targeting SIRTs pharmacologically for the prevention and treatment of neurodegenerative disorders in particular. Fang et al. (2019) discuss the role of SIRTs in the regulation of stem cells.

## Sonic Hedgehog (Shh)

A recurrent finding has been that signaling pathways essential to the brain development are also active in neurogenesis and synaptic plasticity (Yao et al., 2016).

This is demonstrably the case of hedgehog dependent pathways, in particular those dependent upon Shh (Yao et al., 2016). In particular, Shh is active in brain development and also in the maintenance of and proliferation of neural stem cells in the adult brain (Carballo et al., 2018; Schwarz el al., 2012). The action of Shh on neurogenesis appears to be regulated by the primary cilia of hippocampal neurons (Breunig et al., 2008). Shh also participates in axon regeneration (see the bibliometric analysis by Chou et al., 2022). The interplay of the Notch and Shh pathways, where Notch signalling potentiates the cell proliferative effect of Shh has been a recurrent finding in the recent years (Yao et al., 2016). Crosstalk between Shh and Wnt has also been documented (Carballo et al., 2018). Dysfunctions in Shh signalling is present in neurodegenerative diseases, metabolic disorders and depression (Garg et al., 2022; Tayyab et al., 2017; Yao et al., 2016).

## Notch

Increased Notch signalling has been observed following intermittent fasting and exercise (Baik et al., 2020 ; Brandt et al., 2010). During development, Notch-dependant pathways ensure neural stem cells maintenance in the developing brain (Ehm et al., 2010). The Notch signaling pathway also plays an important role in the regulation and plasticity of neural stem cells and neural progenitor cells in several niches of the adult brain (Bagheri-Mohammni, 2022; Brandt et al., 2010; Ehm et al., 2010). Disfunctions of Notch signaling are present in AD and other neurodegenerative diseases (Ables et al., 2011). Notch is a master regulator of neuronal life cycle and synaptic plasticity (Ables et al., 2011). "Notch signaling deeply participates in the development and homeostasis of multiple tissues and organs" (Zhou et al., 2022, p. 1). Transcriptional activation is the most known way Notch activity is enhanced, but protein activation is also possible ('non-canonical' signalling). (Ables et al., 2011). Precursors of Notch receptors are generated by transcription and translation in the endoplasmic reticulum, after which they undergo maturation in the Golgi apparatus (Zhou et al., 2022). Inhibition of GSK-3 by PI3K-Akt increases the activity of Notch. Notch signalling can enhance the activity of the Hedgehog pathway, including its Ssh ligand (Jacobs & Huang, 2020). The hedgehog pathway also regulates Notch signaling (Jacobs & Huang, 2020). Notch activity can be both carcinogenic and anticarinogenic (Zhou et al., 2022).

## Wnt

The Wnt pathways is also important in its neurotrophic activity, including in adult neurogenesis (Arredondo et al., 2020; Alkailani et al., 2022): "[T]he Wnt signaling pathway plays multiple roles in adult hippocampal neurogenesis, including regulation of proliferation, fate-commitment, and maturation of adult-born neurons. " (Arredondo et al., 2022, p. 636). The Wnt/b-catenin pathway is potentiated by AMPK and GSK3β which phosphorilate β-catenin (Hermida et al., 2017; Zhao et al., 2010). Wnt also plays a major role in the differentiation of neuronal precursor cells (Arredondo et al., 2020). General blockade of Wnt almost completely impedes neurogenesis (Arredondo et al., 2020). (Neural) cell proliferation and differentiation which results from autophagy, which is regulated by Wnt (Lorzadeh et al., 2021). Wnt also participates in presynaptic differentiation (e.g., remodeling of the axon and synapse button), presynaptic and post synaptic assembly in the CNS, particularly in the hippocampus, as well as the regulation of the distribution of synapses (Salinas et al., 2013). Wnt also participates in axon regeneration (Chou et al., 2022). While some have mentioned that deficient Wnt signaling is linked with brain tumor genesis (Alkailani et al., 2022), others have found that increased Wnt signaling due to high glucose might be responsible for the link between diabetes and cancer (García-Giménez et al., 2014). Wnt has been reported to activate mTOR through GSK-3 inhibition, which promotes cell growth (Inoki et al.,2006). The association between Wnt signaling and carcinogenesis thus remains to be clarified. Wnt activity is inhibited by Notch and mTOR (Acar et al., 2021; Zeng et al., 2018). A bibliometric analysis of Wnt is provided by Xu et al., (2021).

*Poly (ADP–ribose) Polymerases (PARPs)*

PARPs participate in anti–aging functions such as the promotion of chromosome integrity, metabolic and cell cycle regulation, inflammation, and protein expression and RNA transcription and processing) (Bai, 2015; Cantó et al., 2013). Epigenetic alterations in PARP activities are linked to neurodegenerative diseases and other aging–related diseases due to genomic instability (Amorim et al., 2022; Mao & Zhang, 2022). EIC are modulator of the activity of PARPs (D'Souza et al., 2017; Hofer et al., 2022). Through ADP–ribosylation (the reversible post–translational modification of proteins), the function of the substrate is altered (Li et al., 2022; Lüscher et al., 2021). ADP–ribosylation plays an important role in neurodevelopment, signal transduction, the maintenance of DNA integrity, RNA expression, as well as (neuron) cell fate under conditions of cellular stress (Li et al., 2022; Lüscher et al., 2021). Li et al. (2022) detail the process of ADP–ribosylation and its functioning in regulating physiological and pathological processes, including the roles of different molecules as writers (e.g., NAD+, PARPs), readers (e.g., PARPs, histones) and erasers (e.g., macrodomain–containing proteins) of ADP–ribosylation.

PARPs play a major role in ADP–ribosylation (Cantó et al., 2013; Lüscher et al., 2021). There are 17 PARPs, which are not detailed here (see Bai, 2015), except for the mention that ADP–ribosylation signaling is mainly regulated by PARP1 and to a lesser extent by other PARPs (Cantó et al., 2013; Lüscher et al., 2021). PARPs are activated following DNA damage (Xie et al., 2020). Indeed, the most important function of PARPs is DNA repair, particularly under cellular stress (Cantó et al., 2013).

They are also involved in the regulation of gene expression, immune cell maturation and influence energy metabolism and response to inflammation (Bai, 2015; Cantó et al., 2013). The chronic activity of PARPs can induce NAD+ depletion, which can result in the shutdown of mitochondrial function and glycolytic slowdown through the reduction of NAD+ availability for other processes such as the activation of SIRTs and elevated PARP activation through ROS increase can lead to several diseases affecting major organs (Bai, 2015; Cantó et al., 2013), including increased tumor growth; therefore PARP inhibitors are actively researched (Li et al., 2022).

Table 1. Several signaling proteins involved in NSPAN

| Effector | Type | Main actions |
| --- | --- | --- |
| PI3K | Lipid kinases | Cell proliferation, autophagy, synaptogenesis |
| Akt | Protein kinases | Cell proliferation, apoptosis, insulin signaling, protein synthesis |
| mTOR | Protein kinases | Cell proliferation, protein synthesis, autophagy, immune function |
| AMPK | Protein kinases | Cellular metabolism, nutrient sensing |
| GSK3β | Protein kinases | Glycogen metabolism, cell cycle regulation, cell division, gene transcription, insulin signaling |
| ULK | Protein kinases | Autophagy, mitophagy |
| MAPK | Protein kinases | Cell proliferation and differentiation, immune responses and inflammation, axon arborization |

| PGC-1alpha | Transcription coactivators | Regulation of oxidative stress, inflammation, mitochondrial biogenesis and mitophagy |
|---|---|---|
| NF-κB | Transcription factor | Metabolism, immunity, inflammation, protein synthesis, cell survival, response to tumor microenvironment |
| Sirtuins | Protein deacetylases | Gluconeogenesis, lipid metabolism, mitochondrial respiration, cell cycle control, gene expression, apoptosis, inflammation, insulin secretion, neuroprotection, axonal protection, DNA repair |
| Notch | Transmembrane proteins | Cell fate, development, and differentiation, regulation and plasticity of neural stem cells |
| Sonic hedgehog | Morphogens | Cell proliferation and differentiation, brain development and neural stem cell maintenance and proliferation |
| Wnt | Endoplasmic reticulum proteins | Cell proliferation, differentiation, and survival, differentiation of neuronal precursor cells, presynaptic and postsynaptic assembly |
| PARPs | Nuclear enzymes | Transcriptional regulation, DNA repair, Cell survival and death |

*Note.* This table has been created with the help of GPT-3.5 and its content thoroughly verified

## C) Presentation of the Annotated Bibliography

In the sub-section focusing on exercise, the following literature reviews are summarized: Radak et al. (2007) focused on the role of exercise in regulating ROS in the brain and discussed the implications with regards to neuroprotection; Powers et al. (2020) described exercise as a producer of ROS and a cause for oxidative stress – mentioning the hormesis hypothesis (the experience of mild stress prepares neurons and tissues to more important stressors); Walsh et al. (2020) which focused on the link between exercise, BDNF, as well as cognition; Reddy et al. (2022) which discussed the cellular processes involved in exercise-induced neuroregulation; Chow et al. (2022) focused exerkines' health implications, including cognitive decline; Mattson & Arumugam (2018) discussed epigenetic changes leading to mitochondrial disfunctions (e.g., decreased NAD+:NADH ratio, leading to decreased SIRT signaling) and the alleviating role of exercise and IF; Lucassen et al. (2010) focused on the neuroprotective role of exercise and sleep, as well as the deteriorating effects of stress and inflammation.

In the section on IF, the following literature reviews are summarized. Mattson et al. (2018): focused on the importance of intermittent metabolic switching (involved in both IF and CR) in NSPAN and the neurodegenerative danger of lifestyles without opportunities for intermittent metabolic switching; Seidler & Barrow (2022) examined the role of BDNF in the relation of IF with cognitive function; Cherif et al. (2016) examined IF, CR and Ramadan fasting in relation to cognitive function – mentioning the up-regulation by IF; Wang et al. (2020) focused on the influence of IF on inflammatory markers – which are related to NSPAN.

In the section on CR, the literature reviews that follow are summarized: Almendariz-Palacios et al. (2020 – updated review) : focused on the signaling pathways involved in reduced cellular aging stemming from CR (relevant aspects summarized under this sub-section) and CRMs (relevant aspects summarized in the next sub-section); Ntsapi & Loos (2016): focused on the regulation of autophagy by means of different cycle durations of CR, distinguishing between macroautophagy (starting from 30 minutes of glucose depletion), microautophagy, and CMA (starting from 10 hours of glucose depletion) – CMA as being more targeted than the other forms of autophagy; de Mello et al. (2019); focused on the triumvirate insulin, autophagy, and neurodegeneration – examining the signaling pathways involved; Maharajan et al. (2020): discusses the regulation of stem cells aging and stem cell niches by means of CR – with a focus on signaling pathways; Zhang et al. (2021): focused on the optimization of brain extracellular microenvironment through CR and its role towards neuroprotection; Yu et al. (2020b): examined the restauration and maintenance of cognitive function through CR.

In the section on CRMs, the literature reviews that follow are summarized: Almendariz-Palacios et al. (2020 – summary continues); Bonkowski & Sinclair (2016): explained the use of NAD+ precursors and other SIRTs activators in relation with disease prevention – including neurodegeneration and longevity, with a focus on signaling pathways; Isde et al. (2020): focused on the activation of SIRT1 in particular through phytochemicals, CRMs included; Hofer et al. (2021): examined clinical trials focusing on CRMs and discusses a list of CRMs available through nutrition; Moosavi et al. (2016): focused on the neurotrophic actions of polyphenols, discussing NSPAN; Oluwole et al. (2022) detailed the role of different polyphenol (sub)classes: tannins, phenolic acid, lignans, flavonoids and stilbenes in health including the main outcomes of interest; Rendeiro et al. (2015): examined the effect (direct and indirect) of flavonoids in alleviating neurodegeneration in particular; Zhan et al (2021): focused on the (pre)-clinical uses of resveratrol, addressing neuroprotection through its anti-inflammatory effect; Moradi et al. (2022): focused on the mechanisms of neuronal regeneration following polyphenol intake, addressing ROS scavenging, reduced beta-amyloid accumulation, regulation of apoptosis and gene expression; Grewal et al. (2021): detailed the mechanisms of neuroprotection afforded by quercetin and implications regarding neurodegenerative diseases; Ravula et al. (2021): provided an updated literature review on the prospects for treatment of neurodegenerative disorders offered by fisetin.

In the sub-section on other nutrients inducing neurotrophic effects, summaries of following reviews are provided. Hebergen (2016): examined the role of n-3 polyunsaturated fatty acids and polyphenols in the promotion of adult neurogenesis; Carneiro et al. (2021): provided an updated systematic literature review on the neuroprotective effects of coffee, focusing on several of its compounds; Chen et al., (2010): addressed the protective role of caffeine on the BBB specifically; Cui et al. (2017): focused on the role of vitamin D in brain development, neuroprotection and immunity, distinguishing between genomic and non-genomic actions.

**Discussion**
In the first pages of this contribution (Section A), firstly, the literature on the promotion of NSPAN through EIC and CRMs has been briefly presented. The question of the g-to-k metabolic switch has been addressed and the k-to-g metabolic switching has been mentioned. It was made clear that it is the repetition of these cycles that affords the most health benefits, as "repeating cycles of a metabolic challenge that induces ketosis (fasting and/or exercise) followed by a recovery period (eating, resting and sleeping), may optimize brain function and resilience throughout the lifespan" (Mattson et al. (2018, p. 81).

Thus, in section B, glucose metabolism and expenditure as well as oxidative stress have been discussed, the latter being the result of cellular respiration (electron slippage through the ETC leading increase in ROS). The introduction continued with the review of the main nucleotides (ATP, ADP, AMP), metabolites (NAD+, NADH, BHB), and neurotrophic factors (primarily BDNF), as well as protein kinases, and signaling proteins involved, such as PI3K, Akt, mTOR, AMPK, GSK3β, ULK, MAPK, PGC–1α, NF–κB, and sirtuinsas well as Notch, Shh and Wnt; and selected signaling pathways. While the focus here has been on neurogenesis and neuroprotection, other important functions involved have been discussed (e.g., the regulation of energy metabolism, oxidative stress and inflammation, immune function, apoptosis, autophagy, gene expression and protein synthesis as well as tumor suppression), since beyond their primary role, these functions support the main outcomes of interest. Beckervordersandforth (2017) discuss the role of mitochondrial metabolism in adult neurogenesis. Guo et al. (2022) review the literature on aging in more general terms, related diseases and interventions. Carrasco et al. (2022) and Mittelbrunn & Kroemer (2021) examine the role of T cells in neurodegenerative diseases among others health consequences related to *inflammaging*.

In Section C, a selective list of reviews of interest has been included and a summary of each is provided in the Appendix. The selection of reviews was organized in five sub–sections covering, in relation to neuroprotection and/or neurogenesis: exercise, IF, CR, CRMs as well as other nutrients inducing neurotrophic effects. The aim here was that each sub–section provided overall a general overview of the benefits of its topic (e.g., CR) in relation to neuroprotection and neurogenesis, with each summarized literature review adding specific aspects to this goal.

*Other implications for healthy individuals and policy*

Pharmacological cognitive enhancement (PCE) has not yet been mentioned per se in this contribution. While improving or maintaining NSPAN requires months if not years, PCE usually focuses on the hope of rapid gains and is particularly relied upon by students and pressured professionals (Franke & Lieb, 2013; Mayor et al., 2020). Users of PCE rely upon pharmacological solutions (e.g., methylphenidate, Adderall, or modafinil), generally deviating from the inteded use of these medications, or rely upon illicit drugs. The attitude of the public towards PCE is generally rather negative (Mayor et al., 2019), due to the conception of PCE as focusing on short term performance at school or on the workplace that is considered artificial and therefore unfair and undeserved (Faber et al., 2016). The adoption of NSPAN, with a focus on individual health and collective welfare rather than on achievement might be more acceptable to the public: O'Connor and Joffe (2015) have for instance indicated that individuals are more tolerant of CE when its aim is to compensate for a cognitive deficiency or the maintenance of healthy brain function. This might be even more true in periods of collective challenges, such as the COVID–19 pandemic (Massell et al., 2022; Mayor et al., 2022a, 2022b), as the importance of the greater good might be paramount.

The regulation of PCE is an important consideration for government policy, which aims to tackle the task of balancing benefits and risks, that are not negligible in the case of PCE as usually conceived; i.e., with a focus on neurotransmitters and the short term functioning of brain regions (Outram & Racine, 2011). Advocating in favor of NSPAN through EIC and CRMs might be a good strategy, as the benefits (e.g., reduced health burden for the individual and the collectivity) far surpass those of PCE – while allowing for maintained or improved cognitive functioning in the long run; and their risks are inappreciable when performed within reason: for instance, exercise is a source of ROS (Bloomer et al., 2005) and should therefore not be performed in excess. Calorie restriction should not lead to malnutrition.

*Further studies*

The literature presented in the main text, and the reviews summarized in the annotated bibliography have exhibited the implications of EIC and the potential of CRMs towards the enhancement of NSPAN, including in the context of neurodegenerative diseases. Notwithstanding the focus of basic research on rodents, it is believed that the discovered mechanisms are present in humans as well as they relate to the functioning of the eukaryotic cell, which has largely been conserved in mammals (Mattson et al., 2018). But whether or not adult neurogenesis exists in human remains to be definitely established (Denoth-Lippuner & Jessberger, 2021).

In the majority of cases, studies do not compare the effect of the regulation of different signaling pathways on NSPAN, and more often than not, the effect of exercise, calorie restriction, intermittent fasting, and calorie restriction mimetics are not compared either. It thus remains unclear which intervention is more appropriate to achieve improved NSPAN, or in the best case this is a matter of the judgement of the researcher. Further, the same signaling pathway might operate in opposite directions in relation to NSPAN. For instance, the activation of NF-κB has been linked to neuroprotective and neurodegenerative actions (Capece et al., 2022; Mincheva-Tasheva et al., 2013) and it might be related to the decreased risk of depression through the up-regulation of BDNF, but NF-κB is also related to increased neuroinflammation, a risk factor for depression (Roy et al., 2023). It is also unclear whether the heightened activation of Wnt is beneficial or not overall, as it is involved in both carcinogenesis and carcinoprevention (García-Giménez et al., 2014; Inoki et al., 2006). Another such example is SIRT6, which is elevated in chronically exercised individuals and performs both protective and pejorative actions in relation to cancer (Song et al., 2022; Yang et al., 2020); recommendations with regards to NSPAN should consider which effect is more important (protective / pejorative) with regards to the different pathways regulated by EIC and CRM in establishing minimal and optimal thresholds across modalities.

Studies should investigate dose-response and robust effect sizes in all these areas and compare them across modalities of EIC/CRM and signaling pathways as a prerequisite to the setting of thresholds for recommendations with regards to EIC and CRM. Such elements are very infrequently reported in literature reviews (but see Lucassen et al., 2010) and are absent from the large majority of studies in the field. Examples of exceptions are Walsch et al. (2020), which mention that 30 minutes of exercise are necessary for significant changes in BDNF to occur, whereas, in a study in a related field, Moore et al. (2020) note that blood glucose is affected by single short bouts of exercise in a dose-response manner, with benefits starting after only one minute. It follows the target duration of exercise or IF depends on the parameter one wishes to increase, but the information on such aspects and the reliability of these estimates are frequently lacking in the literature (Li et al., 2023). Further research could also compare, in relation to dose-response and effect sizes, studies focusing on EIC and CRMs to those on ketogenic diets which show promise in the prevention of AD (e.g., Neth et al., 2020).

The comparison of the effect of EIC and CRMs on different signaling pathways and modalities of EIC/CRM should ideally be performed within the same study as to minimize the risk of differences in effects being due to methodological differences between studies. In a review (Sharman et al., 2019) only few human trials focusing on phytocomponents in the treatment of Alzheimer's patients were successful. It should also be noted that the sample size should be sufficient to detect small or moderate improvement with adequate power, for instance a sample size of 128 at least for a two-tailed test of the difference in symptoms between two balanced independent groups with a Cohen $D$ of 0.5 (medium effect size) and a statistical power of 0.8 (calculated with G*Power 3) at a $p$ value of 0.05. Below the minimal required sample size to achieve a relevant power, most studies will fail to detect the desired effect size. Further, the use of meta-analyses is not well implanted in the field; the examples of Pillon et al., 2020 and Wang et al., 2020 were mentioned above. An increased reliance upon meta-analyses could provide precious information on these aspects, should their use not be precluded by the heterogeneity in existing studies or the lack of reporting of effect sizes (Melsen et al., 2014).

An important element should be the determination of minimum and optimal thresholds for the adoption of EIC and CRM modalities in humans (e.g., Kelly et al., 2020; Rendeiro et al., 2015; Yu et al., 2020a), if this is possible, or to find reliable and reproducible means for targeting the ideal quantities to individual needs, which is a hallmark of precision medicine (Wang et al., 2023). Importantly, the needs of members of specific groups should be considered.

In the future, research should clarify whether EIC and CRM can play an important role in *targeted* metabolic reprogramming, which is already successfully used in the treatment of cancer, and might be beneficial with regards to NSPAN (El-Sahli & Wang, 2020; Mela et al., 2020; Pateras et al., 2023; Wang et al., 2022b). This also entails, ideally, setting objectives in realistic expected improvements prior to running such research programs.

Studies relying upon mass-spectrometry and nuclear magnetic resonance have the potential to further highlight the value of EIC and CRMs in terms of NSPAN and might help in identifying additional signaling metabolites and other effectors affected by EIC and CRMs which could be investigated in further detail in a second step, but mass-spectrometry and nuclear magnetic resonance are uncommonly used for related research questions (Baker & Rutter, 2023; Kelly et al., 2020; Kristal et al., 2017; Luan et al., 2017).

Timing is important in signal transduction, as also within the cell processes unfold through time, with uninterrupted successions of events (Koseska & Bastiaens, 2017). An overwhelming majority of studies examine the effect of EIC and CRMs at a single timepoint, blurring the complexity of the orchestration of events. Live cell imaging has potential in further determining distinctions in signaling dynamics in the cross-talk between different pathways (Valls & Esposito, 2022) in starved cells and those with high availability of nutrients.

## Conclusion

This review and annotated bibliography has synthesized recent efforts at understanding the mechanisms of through which EIC and CRMs afford enhanced NSPAN. The mentioned benefits of the different approaches in terms of the enhancement of NSPAN, as well as explanations of the pathways and mechanisms through which this goal can be achieved, have highlighted the potential of EIC and CRMs for a healthier aging.

## Acknowledgements

Table 1 was prepared with the help of GPT-3.5 and thoroughly checked. In addition, GPT-3.5 and GPT-4 have been used to support ideation in the revised version of the manuscript (discussion). Except for some elements of Table 1, no content written by generative pretrained transformers was included in this submission.

**Appendix. Annotated bibliography**

**Exercise**

The role of exercise in the management of ROS is discussed by Radak et al. (2007), with focus on the implications on the CNS. It is showed that oxidative damage of ROS is ultimately linked to neurodegenerative diseases. The protective role of exercise with a focus on neurotrophins and trophic factors are discussed; and among the neurotrophins, some details are provided on BDNF, for which the up–regulation is related to exercise. Through the signaling pathways it can activate, BDNF has a protective role with regards to ROS notably. CREB is mentioned as a downstream effector of BDNF which is linked to an increase in BDNF dependent upon the level of ROS. Hence there is a feedback loop, which is potentiated by exercise.

Powers et al. (2020) focus on exercise as a producer of ROS and a cause of oxidative stress, in the skeletal muscle primarily. The concept of oxidative stress is defined as an imbalance in the pro–oxidant/antioxidant ratio that leads to cellular damage. A short explanation on free radicals is provided. And several molecule that compose the ROS and reactive nitrogen species (NOS) labels are briefly discussed. Enzymatic and nonenzymatic antioxidants are presented as central in the cellular control of ROS. The history of research on exercise–induced oxidative stress and on the production of ROS produced in skeletal muscles during exercise is presented as well as the impact of ROS on muscle performance. Mention is made of the hormesis theory – exercise is presented as a mild stressor that prepares the body to future challenges.

Walsh et al. (2020) focus on the exercise–circulating BDNF–cognition dose–response associations. Physical activity is the most important factors increasing BDNF, of which the neurotrophic functions are mentioned. BDNF and its measurement and concentration are discussed across the life span: 20–30 nanograms per liter with a decrease of 0.5 – 5% per year. Modulators of BDNF, notably, physical activity are presented. Brain aging is then discussed with the mention of mechanisms first, and next the presentation of aging as a result of the presented mechanisms and cognitive decline as a consequence of the latter. The discussion the continues with the involvement of BDNF in aging processes, focusing on the brain and changes in cognitive performance. The association of BDNF with exercise, with mentions a few values of interest are then mentioned. Notably, more than 30 minutes of exercise is necessary to increase serum BDNF and exercise intensity might only be marginally related to the magnitude of the increase in BDNF. Chronic effects of interventions targeting basal BDNF, meta–analyses notably show small effects of exercise on serum BDNF. The next presented elements are the inconclusive link between serum BDNF concentration and cerebral blood flow; that cellular signaling is an important confound in the assessment of the effect of physical activity on BDNF, the lack of research on the acute BDNF response and cognition. Finally, the mode of physical activity, exercise intensity and its duration, as well as work and rest periods, prior training and the BDNF response to exercise in the context of aging are discussed.

Reddy et al. (2023) provide a literature review of the processes involved in exercise-induces neuronal regulation. Research on the physiological response to exercise is synthesized with a focus on the proteins and metabolites, such as exerkines (Irisin, Cathepsin B, interleukin-6 – IL-6, Adiponectin, Osteocacin, Leptin, Fibroblast growth factor-21 – FGF-21), growth factors and signaling proteins, that are regulated by exercise, notably. Irisin in involved in neuroprotection, cognitive improvements and the amelioration of Alzheimer's disease. Cathepsin B is involved in cognitive improvements. IL-6 regulates brain metabolism, notably, and reduces brain inflammation. Adiponectin is involved in dendritic arborization, spinogenesis and neurogenesis notably. Isteocalcin regulates cognition, neuroplasticity, memory and learning. FGF-21 reduces neuroinflammation and improves synaptic plasticity. And leptin is involved in neurogenesis, axonal growth and improved spatial memory notably. Other signaling proteins are mentioned: Orexin A, Musclin, Secreted protein acidic and rich in cysteine (SPARC), Meteorin-like protein (METRNL), BDNF, AMPK, SIRT1, PGC-1$\alpha$, and IGF-1. We have mentioned BDNF, AMPK, SIRT1, PGC-1$\alpha$ and IGF-1 in the main text. Orexin A and SPARC have been related to neurodegenerative diseases. Increased musclin is detrimental to metabolism. Musclin secretion is down-regulated by exercise. METRNL is involved in different functions. Finally, exercise mimetics are mentioned.

In Chow et al. (2022), the health implications of exerkines – substances increased by exercise which are involved in signaling – are presented. Mention is made of the health benefits of exercise are linked to the release of exerkines. Although primarily studied in the skeletal muscle, exerkines can be released by a variety of organs. They might promote health notably by diminishing obesity, reducing risks of cardioavascular disease and cancer, and protecting from cognitive decline. A list of exerkines is provided (e.g., Adiponectin, BDNF, FGF21, IL-6, Lactate, Musclin, Nitric oxide) and mention of the location of their origin is made (e.g., muscle, brain, brown adipose tissue). The most common measures of exerkines, their advantages and disadvantages are discussed. Exerkines play different roles in the cardiometabolic system, such as the improvement of angiogenesis for Angiopoietin I, the optimization of glucose and lipids for adiponectin) in the adipose tissue, the skeletal muscle, bones, the liver and gut, the nervous, immune and endocrine systems. The muscle-brain crosstalk is also discussed in the review, with mention of the role of myokines (exerkines released by the skeletal muscle) in hippocampal neurogenesis in particular. The distinction between acute and chronic exercise is important in that matter, as is kynurenine, a liver-secreted exerkine which can protect the brain from stress.

Mattson & Arumugam (2018) focus on exercise and IF in relation to healthy brain aging. Cognitive decline is presented in the context of aging. It accelerates from 50 years of age and increases odds for neurodegenerative disorders. The physiological age of the brain can be modulated by environmental factors, CR, CRMs, and IF notably. Early to mid-life environmental factors can affect the risk of future neurodegenerative diseases. Traumatic brain injury, and the history of intermittent metabolic switching (e.g., through CR, or IF; beneficial) and chronic positive energy balance (detrimental) are the presented, which is followed by mitochondrial dysfunction (notably on axons and dendrites) in the brain aging process. Mitochondrial dysfunction can result notably in decreased NAD+:NADH ratio, which affects signaling (e.g., SIRT). Different issues are caused by mitochondrial dysfunction (e.g., deficient ETC, increased ROS), which can lead to neurodegenerative disorders and focus on CR as an alleviative process. The processes through which oxidative imbalance can lead to impaired brain health as well as the impaired degradation of damaged proteins are discussed. Next, the discussion continues with dysregulated calcium ion homeostasis in neurons (modulator of neuron function and neuronal networks) and deficient cellular stress response as well as the relevant processes and signaling pathways involved. Dysfunctional neuronal activity is the next topic addressed is the review, followed by impaired DNA repair in aging; then, age-related heightened inflammation and impaired neurogenesis, as well as the limited role of telomere attrition in neurons and cell senescence. A discussion of dysregulation in energy metabolism, with a focus on IRS-1, BHB, and omega-3 fatty acids is presented next. Neurodegenerative disorders are related to aging, and the examples of Alzheimer disease and Parkinson's notably are discussed. Intermittent metabolic challenges and pharmacological interventions are mentioned as means of delaying brain aging, improve cognitive function and neuroprotection. The review then focuses on the processes involved, starting with the depletion of the liver glycogen store and lipolysis and the pathways at play. Examples of pharmacological strategies are provided, for instance ketone esters to improve cognition, nicotinamide riboside to extend lifespan and diminish neurodegeneration, as well as mechanisms of action.

Lucassen et al. (2010) focus on the implications of exercise, sleep and antidepressants, as well as the deteriorating effects of stress and inflammation with regards to neurogenesis. The importance of stress, as an alarm system, for the mobilization of the organism and the deleterious consequences of chronic and uncontrollable stress and severe acute stress are presented. These manifestations of stress can lead to diverse pathologies, notably major depressive disorder, of which the authors mention the hippocampal alterations. Stress affects functions of the hippocampus and its structure (e.g., volume loss; aberrations in synaptic organization). Adult neurogenesis is then discussed, including the areas where it has been established to occur in humans – and points of controversy; and the different steps of this process. How exercise plays an important role in improved neurogenesis is discussed, including the regulation of adult hippocampal neurogenesis by exercise and how it is affected by stress (mentioning the suppressive effect of the latter). Mechanisms for the detrimental effect of stress in relation to neurogenesis are detailed, with mention of concentration of neurotransmitters and neurotrophic factors. Different aspects relative to perinatal stress are presented, as well as sleep disruption as a stressor with ramifications in impaired neurogenesis – but sleep disruption during one single night seems to be unrelated to the process. Hormones and neurotransmitters affected by sleep disruption are presented. Inflammation as a cell stressor is discussed as well as its importance in hypothalamic-pituitary-adrenocortical axis dysregulation, and the consequences regarding adult neurogenesis. Stress also affects the prefrontal cortex (PFC), cytogenesis in this area in particular: Dysregulation of the activity of the PFC can result from impaired cytogenesis. The neurogenic theory is described. This theory links increased risks of depression with impaired neurogenesis following stress exposure. The conclusion addresses reduced long lasting reduced neurogenesis as a cause for depression, and the protective role of voluntary exercise and antidepressants.

## Intermittent fasting (IF)

An evolutionary account of the contribution of repeated metabolic switching to improved neuroplasticity is provided by Mattson et al. (2018) in the form of a comprehensive review. IF is presented as the most frequent repeated metabolic switching paradigm, of which different forms are mentioned (alternate day feeding, daily time-restricted feeding) together with exercise. Intermittent metabolic switching involves the glucose–to–ketone (leading to cellular stress resistance) switch and the ketone–to–glucose (leading to cell growth and plasticity). Organisms evolved to perform well under conditions of scarce resources and that metabolic switching is one way to achieve such result, because of the production of ketone bodies (due to lipolysis), which can be used as cellular fuel when glucose is scarce. The neuronal dynamics induced by repeated metabolic switching are discussed, with mention of the cognitive benefits of exercise and neuroprotection as well as synaptic plasticity. This is followed by the signaling pathways elicited by metabolic switching, focusing on the one hand on neurotransmitters and neurotrophic factors, and on the other hand on the management of cellular stress and mitochondrial biogenesis. Following this, the role of hunger hormone ghrelin is introduced, as well as the neurotrophic factors Insulin–like growth factor I (IGF–1) and fibroblast growth factor 2 (FGF–2) as well as BHB. Finally, repeated metabolic switching is presented as a brain health intervention.

Seidler, K., & Barrow, M. (2022). Intermittent fasting and cognitive performance – Targeting BDNF as potential strategy to optimise brain health. *Frontiers in Neuroendocrinology*, *65*, 100971. https://doi.org/10.1016/j.yfrne.2021.100971

The link between IF and cognition is the focus of Seidler & Barrow (2022). As an introduction, how high caloric diets participates in the development of neurodegenerative diseases is discussed. The role of BDNF as a central molecule in healthy aging and adaptation is presented next. The review links CR and IF to BDNF signaling, with mention of its relevance in relation to neurogenesis, synaptic plasticity, long term potentiation, memory and learning, notably. Then the central role of diminished synaptic plasticity and neurogenesis in neurodegeneration is described. In the systematic review part, research on animals is presented, followed by humans. The findings of the systematic review demonstrate that nutritional challenges fulfill adaptative functions, which involve the BDNF/TrkB signaling pathway; and that short vs long term benefits in cognition can take place depending on the duration of BDNF stimulation.

Cherif, A., Roelands, B., Meeusen, R., & Chamari, K. (2016). Effects of intermittent fasting, caloric restriction, and Ramadan intermittent fasting on cognitive performance at rest and during exercise in adults. *Sports Medicine, 46*, 35–47. https://doi.org/10.1007/s40279-015-0408-6

The influence of IF, Ramadan fasting and CR on adult cognitive performance during exercise or at rest is the focus of Cherif et al. (2016). The health benefits of exercise and how it affects the brain, the importance of balanced energy intake and the distinction between calories restriction and IF are mentioned, which is followed by a brief review of the literature on fasting and cognitive function as well as the importance of nutrition for athletic performance. Next CR is presented, with mention of caloric deficit values and benefits associated with CR (e.g., increased lifespan) with accompanying experimental evidence, as which mechanisms for the benefits of CR on cognitive function are presented. The IF is distinguished from religious fasting and alternate day fasting. The cognitive benefits of IF, and the beneficial changes in neurochemistry and neuronal networks that stem from it, including BDNF concentration are discussed. Mention is made of the increase in parasympathetic activity that exercise promotes and the benefits of metabolic switching. Next, the discussion of Ramadan fasting defines its scope and describe its effects (positive and negative) on cognition and other areas (e.g., sleep). A general overview the effects of the diets of interest on cognitive function is proposed, starting with the importance of the relationship between immunity and cognitive function (focusing on cytokines) which is related to interferences of inflammation with in the expression of BDNF and BDNF signaling. After this, the effects of the diets on glucose and lipid metabolism is related to cognition (poorer with hypoglycemia). The importance of lactate (the by–product of glycolysis) for long term memory and of metabolic changes associated with the diets in relation to neuroprotection is discussed. This is followed by the detrimental effects of dehydration on cognition – which start with 1% loss in body weight and the review of evidence relating the combination of exercise and the diets to the healthy brain.

Wang, X., Yang, Q., Liao, Q., Li, M., Zhang, P., Santos, H. O., ... & Abshirini, M. (2020). Effects of intermittent fasting diets on plasma concentrations of inflammatory biomarkers: a systematic review and meta-analysis of randomized controlled trials. *Nutrition, 79*, 110974. doi: 10.1016/j.nut.2020.110974

A meta-analysis focusing on IF and inflammatory biomarkers is proposed by Wang et al. (2020). It is introduced with the mention of the importance of a balanced energy intake in longevity and the risks associated with an over consumption of calories in related to metabolic disorders. The aging processes can be attenuated by IF and energy-restricted diets (variations are discussed). Next energy restricted diets and IF are presented in relation to the activation of AMPK – a central process in the benefits of these diets. The meta-analysis ($k$ = 18, 15 articles). examines the effects of the diets on the concentration of C-reactive protein (CRP), tumor necrosis factor – α (TNF- α) and interleukin-6 (IL-6). Findings show the diets compared with ad libitum feeding affected CRP (IF more than energy restricted diets), but not TNF- α nor IL-6. No evidence of publication bias is reported.

**Calorie restriction (CR)**

Almendariz-Palacios, C., Mousseau, D. D., Eskiw, C. H., & Gillespie, Z. E. (2020). Still living better through chemistry: An update on caloric restriction and caloric restriction mimetics as tools to promote health and lifespan. *International Journal of Molecular Sciences, 21,* 9220. doi:10.3390/ijms21239220

An updated review of the literature (see Gillespie et al., 2016 for the original review) on healthy lifespan improvement in relation to CR and CRMs is performed by Almendariz-Palacios et al. (2020). It starts with the mention of the impact of aging on pathologies, including neurodegenerative disorders, and the role of excessive calorie consumption. Next, the role of CR in the limitation of the effects of aging on the organism and reduced cellular aging is discussed: Aging is related to increased senescence in cells, leading to "key alterations (…) [which] include increased genomic instability, telomere shortening, changes in the epigenome, loss of protein homeostasis (proteostasis), deregulated nutrient sensing, mitochondrial dysfunction, stem cell exhaustion, and altered intracellular signaling" (p. 2). Mention is made of the importance of ROS in mitochondrial dysfunction, leading to cellular senescence and the role of CR on minimizing such consequences of aging is presented. The mention of the hormesis theory (preparative role of mild cellular stress leading to protection) as an explanation for the effectiveness of CR follows. The downstream effects of CR, including cellular nutrient sensing (and the related pathways, e.g., AMPK up-regulation through increased AMP/ATP ratio; mTOR down-regulation in the case of calorie restriction) are discussed as an important factor in healthy lifespan optimization. Following this, the role of amino acid restriction in healthier lifespan is presented, in relation to mTOR amino acid sensors, notably sestrin2 and cytosolic arginine sensor for mTORC1 (CASTOR), which inhibit mTOR1 in case of arginine and leucine depletion. The effects of CR and amino acid restriction are compared (similarities and differences), in terms of the involved signaling pathways. The authors then discuss CRMs (see below).

Ntsapi, C., & Loos, B. (2016). Caloric restriction and the precision-control of autophagy: A strategy for delaying neurodegenerative disease progression. *Experimental Gerontology, 83*, 97-111. https://doi.org/10.1016/j.exger.2016.07.014

Ntsapi & Loos (2016) focus on autophagy and CR. First, estimates of people aged 60 and more by 2050 (22% worldwide) and related challenges are mentioned, notably in relation to chronic disease; including neurodegenerative diseases, linked to the accumulation of proteins. Autophagy is presented as a protective means in this respect and in reducing ROS. Autophagy can be enhanced through CR, of which the general health benefits, such as increased longevity as mentioned. Autophagy is presented as a response to nutrient deprivation with implications in neuronal homeostasis and cell survival. Three types of autophagy exist in mammals: "macroautophagy, chaperone-mediated autophagy (CMA), and microautophagy" (p. 98), which are defined and characterized through their signaling pathways. The diminishing of the autophagy types with aging is related to different processes, which can be counteracted through CR. For instance, macroautophagy is increased after already 30 minutes of nutrient depletion, whereas CMA requires 10 hours of caloric intake abstinence after depletion. CMA is preferable to macroautophagy, which is less selective in proteins degradation and could be counterproductive. The neuroprotective role of autophagy is then presented, and a distinction between macroautophagy and CMA in terms of outcomes is made. Most studies in relation to neurodegenerative diseases were interested in macroautophagy. The association of macroautophagy and CMA with different neurodegenerative diseases is discussed as well as the detrimental role of specific aggregative proteins, and the importance of mitochondrial dysfunction (which increases with age) in insufficient clearance of these proteins. The review ends with the description of the implication of targeted autophagy increased healthy lifespan, presenting IGF-1 as a central aspect in the effectiveness of calorie restriction in this process.

de Mello, N. P., Orellana, A. M., Mazucanti, C. H., de Morais Lima, G., Scavone, C., & Kawamoto, E. M. (2019). Insulin and autophagy in neurodegeneration. *Frontiers in Neuroscience*, *13*, 491. https://doi.org/10.3389/fnins.2019.00491

de Mello et al. (2019) focus on insulin and autophagy in relation to neurodegeneration. That metabolic disorders are among the most important cause of reduced life expectancy and comorbidity, including neurodegenerative diseases, is discussed first and linked to imbalanced insulin signaling, which plays an important role in neurodegeneration. How different processes such as autophagy and apoptosis, which rely upon different signaling pathways (e.g., PI3K, AMPK, Akt, mTOR), can alleviate or pejorate this is examined next. The "review summarizes the origin and the role of insulin in the CNS, and discusses the relationship between insulin and autophagy in some neurodegenerative diseases" (p. 2). Insulin resistance is presented as a factor of neurodegeneration. The effects of insulin, including on neurons are presented and brain insulin is linked to IGF-1; and insulin signaling and autophagy to neurodegeneration through different signaling pathways and several neurodegenerative diseases. CR and exercise are finally discussed as important means of regulation of autophagy.

Maharajan, N., Vijayakumar, K., Jang, C. H., & Cho, G.-W. (2020). Caloric restriction maintains stem cells through niche and regulates stem cell aging. *Journal of Molecular Medicine*, *98*, 25-37. https://doi.org/10.1007/s00109-019-01846-1

Maharajan et al. (2020) focus on the optimization of stem cells function through CR. First, mention is made of the importance of stem cells in the aging process and homeostasis in general. The review continues with mention of the presence of stem cells in niches which communicate with their environment; and next that calorie restriction and CRMs as beneficial in terms of health and longevity. Stem cell aging is defined as a decrease in their functions and list different associated impairments, which is discussed in relation to their intrinsic and extrinsic influences. The components of the niche are defined as: "cells that are physically attached to the stem cells (stromal cells and mesenchymal cells), adhesion molecules on the membrane, and membrane bound ligands or receptors in the niche cells." (p. 28) as well as "secretory or soluble factors, such as chemokines, hormones, GF, Hedgehog, TNF, nuclear factor-kappa B (NF-kB), Wnt3, EGF, and Notch, which are produced from stem cells, stem cell progenitor cells, or niche cells." (p. 29). How the niche affects the stem cell, for instance how the alteration of some components of the niche is detrimental to the stem cell, is discussed with examples in different organs, notably the brain. Then the regulation of the niche by non-cellular components is presented: the extracellular matrix and the importance of integrins, which notably promote the stem cells adhesion to the extracellular matrix, of which deteriorated signaling is involved in neurodegeneration. Following this, how aging is affected by energy metabolism and CR, through the activation or inhibition of different pathways (e.g., AMPK, SIRT, mTOR) is presented with the mention of normal caloric intake in relation to lower self-renewal of stem cells and anomalous functioning, but higher differentiation of stem cells; and opposite effects of CR. This differentiated impact of the presence or absence of abundant nutrients and the pathways involved are discussed. CR and CRMs are finally presented as a means to increase healthy lifespan notably through improved stem cell niche and stem cell proliferation.

Zhang et al. (2021) examine the optimization of brain extracellular microenvironment through CR, with a focus on minimizing brain aging and symptoms of neurodegenerative diseases. First, mention is made of the consequences of aging at the cellular level (e.g., cellular senescence, with disrupted gene expression) and in the brain (e.g., neuronal damage). Then the importance of the extracellular matrix and cerebrospinal fluid to the health of neurons and its broader consequences for the organism (e.g., neurodegeneration) is presented, followed by review of the signaling pathways linked with cerebral inflammation and the role of cytokines (inflammation) and microglial cells (sensing and cytokine secretion notably) in this process. The detrimental effect of metabolic waste in the extracellular microenvironment and the reduction of antioxidants that accompanies aging are presented, as well as the clearing of waste through cellular autophagy and the transport of waste outside the brain through the BBB. Next untimely removal of metabolic waste in aging due to the disruption of the permeability of the BBB are discussed. The subsequent topics addressed are the modulation of gene expression which is impacted by aging through alterations of chromosome structures; and how the cellular micro-environment can be enhanced through CR, in relation to diminished risks of neurodegenerative diseases. Implications of CR with respect to the cellular microenvironment at the molecular level (e.g., inhibition of mTOR, activation of SIRT, up-regulation of CREB leading to the increase in BDNF) are examined before the presentation of the increase in neurotrophic factor concentration as a consequence of CR and the role of CR in improving the micro-environment at the cellular level, distinguishing between mitochondrial biogenesis and autophagy, and gut microbiota. Implications of CR at the tissue level are provided, with mention of the reduction of tissue inflammation through CR. The influence of CR on the permeability of the BBB, synaptic plasticity, neurogenesis and neuron immunity are then addressed. CR interventions are the focus of the next section: a reduction in aging is induced by CR due to neurogenesis. Implications for specific diseases such as Parkinson's and Huntington are discussed.

Yu and colleagues (2020b) discuss the restauration and maintenance of cognitive function through CR. Six dimensions of cognitive abilities decline with aging. The social and economic costs of such decline as well as its pharmacological treatment are presented. CR is discussed as an alternative method of dealing with cognitive decline. Next, the amelioration of a variety of aging-related health conditions through CR is detailed, and the potential mechanisms relating each time CR to cognitive function are presented: oxidative stress (areas related to cognition are more at risk of ROS damage), inflammation (secretion of inflammatory molecules by senescent cells, and their proinflammatory environment), neurogenesis and synaptic plasticity (neurotrophic factors and BDNF signaling), and neuroprotection (preservation of gray and white matter). In conclusion, a call is made for further studies to investigate several aspects of CR in order to optimize its benefits in relation to the duration of CR, the amount of CR, and the combination of CR with other interventions.

**Calorie restriction mimetics (CRMs)**

The summary of Almendariz-Palacios et al. (2020) is now continued in relation to CRMs. CRMs are portrayed as a means to promote healthy lifespan: Rapamycin is an immunosuppressant and antitumor medication, related to decreased mTOR1 signaling. Mention is made of its benefits in terms of lifespan extension and reduced risk of cancer as well as the augmentation of autophagy and neuroprotection among other benefits. Rapamycin analogs are then discussed. Resveratrol exerts its effects through AMPK and SIRT activation, but can have detrimental effects in high concentration. Metformin, a type 2 diabetes medication, is also discussed along with the pathways it regulates.

Bonkowski & Sinclair (2016) focus on NAD+ precursors and other SIRT activators. First, the importance of disease prevention – including neurodegeneration, and implications for longevity extension are presented. Several signaling pathways conserved from simple to complex organisms are examined next. These explain part of the effectiveness of calorie restriction in healthy aging, including AMPK, SIRTs and mTOR regulation. Then, natural molecules that are capable of activating the mentioned pathways are discussed. The review focuses on SIRTs and their activators (STACs), including NAD+ precursors. Animal models, primate and human studies are discussed in the presentation of the mechanisms of STACs. Following this, the specifics of NAD+ precursors are discussed, as NAD+ is required for the activity of SIRTs. "NAD-boosting molecules constitute a newer class of STACs gaining attention as a way to restore NAD+ levels in elderly individuals and potentially activate all seven sirtuins with a single compound. " (Bonkowski & Sinclair, 2016, p. 17).

Iside et al. (2020) focus on the activation of SIRT1 by phytochemicals. The epigenic modifications associated with aging are presented before the role of SIRTs in reducing such consequences, and in improving a range of parameters that relate to better health, including neuroprotection. The review continues with the importance of diet in health, including the protective nature of phytochemicals, like "polyphenolic substances such as resveratrol, quercetin, curcumin, and fisetin, and of natural non-polyphenolic substances such as berberine" (p. 2). These molecules are related to the expression and activity of SIRTs. Studies which exemplify the activation of SIRTs by resveratrol, quercetin, berberine, curcumin and fisetin are reviewed, focusing on the signaling pathways involved. The conclusion focuses on the activation of SIRTs through natural CRMs, which might explain the role of these molecules in preserving or improving health, and the bioavailability of the mentioned phytochemicals.

Hofer and colleagues (2021) focus on the availability of CRMs in nutrition. Although nutrition is essential in health, there is a relative lack of agreement regarding the definition of a healthy diet. CR is presented as a point of consensus in this debate. CRMs are defined as "pharmacologically active substances that mimic some of CR's myriads of effects. At the core of the CRM definition, we and others argue that potential CR-mimicking compounds should in principle increase life- and/or healthspan and ameliorate age-associated diseases in model organisms, thus often the simultaneous use of the term 'anti-aging substances' " (p. 2). A table of the sources of nutritional CRMs is provided, with several classes: glycolysis inhibitors, hydroxycitric acid, NAD+ precursors, polyamines, polyphenols, salicylic acid. Example of studies for each of these classes of CRMs are examined. The conclusion elaborates on the need for further studies examining points of uncertainty and confirming the beneficial effects of some CRMs in humans.

Moosavi et al. (2016) focus on the neurotrophic actions of polyphenols. First, the sources and health protective role of polyphenols in relation to different diseases – including neurodegenerative ones is presented, distinguishing between different neurotrophic functions of polyphenols (e.g., neuroprotection, neuronal proliferation, anti-oxidant effect). Details are provided on: the signaling pathways (e.g., Trk, mTOR, ERK, PI3K, CREB, Nrf2) that are targeted by different polyphenols (e.g., curcumin, quercetin, EGCG, carnosic acid) and the neurotrophic factors that are affected (BDNF, nerve growth factor – NGF, glial cell line–derived neurotrophic factor), and the functions that are achieved (e.g., antidepressant-like effects, cognition enhancement, attenuation of neuronal degeneration).

Oluwole et al. (2022) focuses on the role of different polyphenol (sub)classes: tannins, phenolic acid, lignans, flavonoids and stilbenes. The classification of the large variety of polyphenols (flavonoid vs non-flavonoid) depends on the quantity of phenols their molecule contains. The functions and uses of the mentioned (sub)classes are summarized. Phenolic acids are defined, and their biological functions are mentioned: antioxidant (e.g., ROS scavenging), gastroprotective, antidiabetic, cardioprotective, chemo-preventive (e.g., lipid and protein metabolism), antimicrobial, anti-inflammatory. Their role in the prevention of neurodegeneration and neuronal injury is detailed. Next, tannins are presented, with their antimicrobial, antiviral, anti-mutagenic properties as well as their role in diabetes. Stilbenes are discussed, with mention of their chemo-preventive properties (e.g., antioxidant and anti-inflammatory) and their role in obesity and diabetes as well as the prevention of neurodegeneration. After this, the discussion continues with lignans, with chemo-preventive (e.g., hormone regulation, inhibition of cell growth), cardioprotective, antimicrobial properties. Finally, flavonoids are defined their biological functions (e.g., antioxidant, regulation of hydroperoxides), with mention of their properties: inhibition of lipid peroxidation, chemo-prevention (e.g., protective role against cancer), antimicrobial and anti-viral protection, cardioprotection, hepatoprotection, anti-inflammation and anti-neurodegeneration.

Rendeiro et al., (2015) examine the mechanisms of action of flavonoid polyphenols in the context of the prevention of neurodegenerative diseases. The elevation of serum nitric oxide is one of the benefits of flavonoids. The bioavailability of flavonoids metabolites in the brain is mentioned in relation to their ability to cross the BBB. Several flavonoid metabolites have been detected in the brain, which supports such requirement for neuroprotection. Human cognition is influenced by flavonoids and reported benefits of several groups of flavonoids are mentioned. The mechanisms of action of flavonoids, i.e., the modulation of synaptic plasticity and the pathways of such action are discussed, including the indirect mechanisms which relate to the improvement of the neurovascular system and cerebrovascular functioning (e.g., blood oxygenation levels, increased brain vascularization), through increased nitric oxide bioavailability notably. The correlational nature of the reported association is discussed with the need for research that improves the mechanistical assumptions relying upon strong empirical evidence, some of which is already available. That neither long term effects of flavonoids have yet been thoroughly investigated in humans, nor their recommended daily dose is also discussed.

Zhang, L.-X., Li, C.-X., Kakar, M. U., Khan, M. S., Wu, P.-F., Amir, R. M., Dai, D.-F., Naveed, M., Li, Q.-Y., Saeed, M., Shen, J.-Q., Rajput, S. A., & Li, J.-H. (2021). Resveratrol (RV): A pharmacological review and call for further research. *Biomedicine & Pharmacotherapy*, *143*, 112164. https://doi.org/10.1016/j.biopha.2021.112164

    Zhang et al. (2021) focus on the pharmacological actions of resveratrol, a stilbenoid polyphenol. First a brief presentation of resveratrol is proposed with a list of its functions (e.g., cardioprotection, reduction of neurotoxicity) and a brief mention of the signaling pathways involved (e.g., reduction of oxidative stress, improvement of mitochondrial health). Next, the chemical structure of resveratrol is presented, with mention of most important nutritional sources (e.g., tomatoes, peanuts, red grapes, different berries, pomegranates, cranberries, dark chocolate) and intake recommendations. The presentation continues with the pharmacokinetics of resveratrol including its metabolization in the liver, its plasma concentration and its distribution in different organs. Its mechanisms of action are explained, focusing in more detail on the signaling pathways of which it supports the regulation. Finally, a health perspective on resveratrol is proposed in the form of a synthesis, focusing on its antioxidant and anti-inflammatory actions as well as its role in protection against neurodegenerative diseases among others.

Moradi, S. Z., Jalili, F., Farhadian, N., Joshi, T., Wang, M., Zou, L., Cao, H., Farzaei, M. H., & Xiao, J. (2022). Polyphenols and neurodegenerative diseases: Focus on neuronal regeneration. *Critical Reviews in Food Science and Nutrition*, *62*(13), 3421–3436. https://doi.org/10.1080/10408398.2020.1865870

    Moradi et al. (2022) focus on neuronal regeneration following polyphenol intake. Following the mention of the scarcity of organ donation and the importance of regenerative medicine, the review continues with explanations of nervous system diseases (such as AD, PD, HD, ALS) and their therapy, notably focusing on cytokines, growth factors, neurotrophic factors and stem cells. Next, the therapeutic effects of polyphenols in neurodegenerative disorders are examined, distinguishing between flavonoids, isoflavones, phenolic acids and tannins as classes of polyphenols for which such benefits were investigated. The mechanisms of action are detailed, such as ROS scavenging, reduced beta–amyloid accumulation, regulation of apoptosis and gene expression. Example of studies on the regenerative effect of polyphenols are proposed, focusing on a single or multiple molecules. Then the role of stem cells in neuronal regeneration is examined: "Stem cell therapy is one of the most important branches of regeneration medicine that has enhanced the researcher's hope to treat or improve the condition of the patient with varying debilitating diseases" (p. 4). Examples of studies are provided. Neuronal regeneration through chemical stimulation and the emerging role of gene therapy are presented. A neuronal perspective then addresses the role of polyphenols in regeneration. The antioxidative properties of polyphenols, the potential improvements in neurogenesis, neuronal connectivity, nerve damage, neuron survival notably stemming from of a variety of polyphenols are presented next. Finally, challenges and future perspectives in the field are discussed and the potential of the mentioned treatments in alleviating consequences of aging is highlighted.

Grewal, A. K., Singh, T. G., Sharma, D., Sharma, V., Singh, M., Rahman, Md. H., Najda, A., Walasek-Janusz, M., Kamel, M., Albadrani, G. M., Akhtar, M. F., Saleem, A., & Abdel-Daim, M. M. (2021). Mechanistic insights and perspectives involved in neuroprotective action of quercetin. *Biomedicine & Pharmacotherapy*, *140*, 111729. https://doi.org/10.1016/j.biopha.2021.111729

Grewal et al. (2021) delineate the mechanisms of neuroprotection of quercetin, a flavonoid polyphenol. The neuroprotective effect of polyphenols (notably reduced oxidative damage and reduction in risks for several neurodegenerative diseases) is examined. Nutritional sources of quercetin (e.g., red onion, garlic, apple, pomegranate), its chemical formula and pharmacological action are presented. The neuroprotective effect of quercetin can be enhanced through nano-emulsion. The pathogenesis of neurodegenerative diseases is discussed. It notably involves mitochondrial dysfunction, oxidative stress, elevated apoptosis, and inflammation. The review elaborates on different neurodegenerative diseases for which quercetin has been shown to reduce risks and symptoms and proposed mechanisms of action (e.g., for Alzheimer's disease: improvement in neuroinflammation, oxidative stress, amyloid plaques cholinesterases, notably; for Parkinson Disease: improvement in neuroinflammation, oxidative stress, apoptosis, autophagy, notably; for Huntington disease, improvement in neuroinflammation, oxidative stress, cognitive deficits, neuronal dysfunction, notably). Finally, the signaling pathways influenced by quercetin are detailed, among others: paraoxonase 2 (PON2), the nuclear factor erythroid 2-related factor 2 (Nrf2) - Adenylate/uridylate-rich elements (ARE).

Ravula, A. R., Teegala, S. B., Kalakotla, S., Pasangulapati, J. P., Perumal, V., & Boyina, H. K. (2021). Fisetin, potential flavonoid with multifarious targets for treating neurological disorders: An updated review. *European Journal of Pharmacology*, *910*, 174492. https://doi.org/10.1016/j.ejphar.2021.174492

Ravula et al. (2021) provide an updated review on the neuroprotective and neurogenerative impact of the flavonoid polyphenol fisetin. Neurological diseases are a leading cause of disability and mortality and the difficulty of therapy of such diseases is discussed. The complementary role of flavonoids in the treatment of neurodegenerative disorders is presented with mention of the "potential of flavonoids as antioxidants, antiviral, anti-inflammatory, anti-carcinogenic, anti-bacterial, neurotrophic, neuroprotective, and immune-stimulants" (p. 2). The functions of fisetin in particular in the prevention of neurodegeneration are mentioned: reduced neuroinflammation, improved immune response, and the optimization of signaling pathways that are disrupted by aging and in neurodegenerative disorders. Research showing that fisetin is effective in improving cognition is presented, which is followed by the mechanisms affected by fisetin, some of which were not discussed in the main text of our review, such as: the 'modulation of Cdk5/p35' (associated with neuroinflammation, synaptic damage notably and neuronal death), the 'regulation of eicosanoids' (pro-inflammatory lipid componds), 'saving ATPase' (important in homeostasis, impaired by oxidative stress), 'fisetin-TFEB—MTORC1-Nrf2 linkage' (role in autophagy notably), the 'modulation of Kelch-like ECH-associated protein 1 (Keap 1)- nuclear factor erythroid 2-related factor 2 (Nrf2)- antioxidant response elements (ARE) pathway' (control of xenobiotic damage and oxidative stress; maintenance of enegery metabolism and redox balance in the cells), 'Regulation of advanced glycation end products (AGEs)' (inhibit protein functions, impair antioxidant enzymes, aggregate proteins, notably) ; as well as the 'modulation of CREB' (involved in learning and memory), 'NAD+ degradation' (see above), the 'restauration of synaptic proteins', the 'elevation of Acetyl CoA' (involved in glycolysis and fatty acid synthesis & oxidation, abnormalities in Acetyl-CoA can lead to dementia, precursor to neurotransmitter Acetylcholine), and the 'modulation of NF-kB' (pro-inflammatory, regulates cytokine expression, involved in neurodegeneration). Research on fisetin in relation to neurodegeneration is summarized in Table 1. Figure 7 indicates that fisetin decreases or inhibits among others: lipid peroxides, IL-6, TNF-alpha, NF-kB, nitric oxide, superoxides, serum homocysteine; and increases or activates among others: BDNF, acetylcholine, butyrylcholine, serotonine, noradrenaline, Nrf2, TrkB, MEK-ERK.

**Other nutrients inducing neurotrophic effects**

Several molecules which are not considered CRMs have neurotrophic effects. Some of these molecules are briefly introduced in the continuation of this annotated bibliography.

Heberden, C. (2016). Modulating adult neurogenesis through dietary interventions. *Nutrition Research Reviews*, *29*, 163-171.

> The modulation of adult neurogenesis with polyphenols and n–3 polyunsaturated fatty acids is examined in Heberden et al. (2016). The potential of the human brain to regenerate might be less than that of rodents and limited to specific areas with neural stem cells, which are presented: the subgranular zone of the dentate gyrus, the subventricular zone of the lateral ventricle, and the median eminence of the hypothalamus. The regulators and effectors of neurogenesis (through neurotrophins and growth factors) are examined with mention of the concentrated nature of neural stem cells niches. The role of leptin as a promoter of neural stem cell proliferation, as well as other peptide hormones, is highlighted. The detrimental effect of aging on neurogenesis is examined, following with dietary influences on this process. The unfavorable effects of overnutrition are addressed, notably in terms of proliferation and survival of new neurons. The potential of dietary intervention through polyphenols (see also above) and n–3 polyunsaturated fatty acids is addressed: beneficial in terms of neural stem cell proliferation and differentiation, neurogenesis, and arborization of new neurons. The targets and signaling pathways involved are detailed, such as AMPK, BDNF/CREB, NF–kB, SIRT1. The gut microbiota is presented as a potential effector of nutrition induced adult neurogenesis, for instance through their action on metabolism and inflammation.

Carneiro, S. M., Oliveira, M. B. P., & Alves, R. C. (2021). Neuroprotective properties of coffee: An update. *Trends in Food Science & Technology, 113,* 167–179. https://doi.org/10.1016/j.tifs.2021.04.052

> Carneiro et al. (2021) provide an updated systematic literature review on the neuroprotective effects of coffee. The stimulant properties in the CNS are presented as a factor contributing to the large worldwide consumption of coffee. The health benefits of coffee are also mentioned. The aims of the review are to synthesize the literature relating coffee consumption to the reduction of the risk of neurodegenerative disorders and the mechanisms involved. There are more than 1,000 components in coffee of which the most important are presented. The most studied, caffeine is antagonist of the inhibitory neurotransmitter adenosine, which notably explains the stimulant properties of coffee. Another important compound is trigonelline, which is related to neuroprotection. The neurotransmitter serotonin is also a compound of coffee. Coffee also contains polysaccharides, which have an impact on the gut microbiome that enhances cognitive function and the anti–oxidants cafestol and kahweol. Nicotinic acid (B3) is a component of roasted coffee. Other components are enumerated and their health benefits mentioned (e.g., antioxidant, anti–inflammatory, anticarcinogenic). Next, the role of coffee in the prevention of several neurodegenerative diseases is examined: for instance, through the reduction of excitotoxicity, oxidation and inflammation (caffeine), or the prevention of memory impairments by trigonelline. The phenolic components and other components of coffee and their neuroprotective properties are also mentioned and research on the protective role of coffee in the risk for different neurodegenerative diseases is detailed. The positive global health impact of coffee, notably in relation to neurodegenerative disorders, is discussed.

Chen, X., Ghribi, O., & Geiger, J. D. (2010). Caffeine protects against disruptions of the blood–brain barrier in animal models of Alzheimer's and Parkinson's diseases. *Journal of Alzheimer's Disease*, *20*(s1), S127-S141. https://doi.org/10.3233/JAD-2010-1376

> Chen et al. (2010) focus on the protective role of caffeine with regards to the integrity of the BBB, which is one of the neuroprotective effects of caffeine. The function of the BBB (to isolate the brain from toxins, notably) and the structure of the BBB are first described. The disruption of the BBB is mentioned as one of the causes of the onset of AD

(notably as a result of high cholesterol levels), and potentially PD (BBB disruption occurring before the loss of dopaminergic neurons, characteristic of PD). Three mechanisms through which caffeine is believed to protect the BBB are examined: notably by "blocking cell surface adenosine receptors, through inhibition of cAMP phosphodiesterase (PDE) activity, and by affecting the release of calcium from intracellular stores" (p. S131). Next, the integrity of the BBB through chronic caffeine consumption might explain part of the neuroprotective effect of caffeine, and examples relating for instance to better memory and cognition, reduced loss of dopaminergic neurons, as indicators of such effect are provided. The protective actions of caffeine on the BBB are mentioned as primary in comparison to neuroprotective actions of caffeine.

Cui, X., Gooch, H., Petty, A., McGrath, J. J., & Eyles, D. (2017). Vitamin D and the brain: Genomic and non-genomic actions. *Molecular and Cellular Endocrinology, 453,* 131–143. https://doi.org/10.1016/j.mce.2017.05.035

Cui et al. (2017) examine the role of vitamin D in brain development, neuroprotection and immunity. Vitamin D is presented as a developmental neurosteroid, positively involved in brain development and neuroprotection. The signaling functions of vitamin D through the Vitamin D Receptor (VDR) are detailed. Brain morphology and physiology are importantly linked to vitamin D. The goals of the review are presented: the examination of the actions of vitamin D in "brain cell differentiation, neurotransmitter release, and calcium signaling via its genomic and non-genomic functions." (p. 132). VDR is distributed in the brain. Rodent models are discussed first as ("[the] pattern of VDR distribution indicates that vitamin D may be involved in the proliferation and/or differentiation of neuronal stem cells", p. 132), then the human brain as ("[the] VDR protein was also identified in the human brain and the distribution pattern of the VDR was found to be strikingly similar to that reported in rodents, p. 133). The genomic actions of vitamin D are addressed, in particular its involvement in apoptosis, cell proliferation, and neural growth. The role of vitamin D in the regulation of the development of dopaminergic neurons and other cells, and then the differentiation of adult neural stem cells and myelination is also examined. The effect of vitamin D on neuronal survival, programmed glioma cell death and neurotransmitter release is addressed. Which is followed by the neuroprotective effects of vitamin D in the aged brain and its relation to neurotrophic factors, BDNF notably. Next, mention is made that vitamin D can compensate neuroinflammation in the aging brain, as well as immune response. Also, "vitamin D could potentially increase local estrogen synthesis in glial cells, which is important to maintain neuronal function" (p. 137). The non-genomic functions of vitamin D, an underexplored area, are then examined, this includes: the role of vitamin D in the regulation of calcium and kinase activated pathway signaling recently discovered in the developing cortex, and other modulatory effects of vitamin D, which for some have neuroprotective properties. The therapeutic potential of vitamin D, and the negative consequences of deficiencies in vitamin D are finally discussed.